\documentclass[prc,aps,showpacs]{revtex4}
\usepackage{graphicx}
\begin{document}

\title{Dispersion Relation Constrained Partial Wave 
       Analysis \\ 
       of $\pi$N Elastic and $\pi N\to\eta N$ Scattering 
       Data: The Baryon Spectrum}

\author{R.~A.~Arndt}
\email{arndt@reo.ntelos.net}
\author{W.~J.~Briscoe}
\email{briscoe@gwu.edu}
\author{I.~I.~Strakovsky}
\email{igor@gwu.edu}
\author{R.~L.~Workman}
\email{rworkman@gwu.edu}
\affiliation{Center for Nuclear Studies, Department of
        Physics, \\
        The George Washington University, Washington,
        D.C. 20052, U.S.A.}
\author{M.~M.~Pavan}
\email{marcello.pavan@triumf.ca}
\affiliation{TRIUMF, Vancouver, B.C. V6T~2A3, Canada}

\begin{abstract}
We present results from a comprehensive partial-wave analysis 
of $\pi^\pm$p elastic scattering and charge-exchange data, 
covering the region from threshold to 2.1~GeV in the lab pion 
kinetic energy, employing a coupled-channel formalism to 
simultaneously fit $\pi^-p\to\eta n$ data to 0.8~GeV.  Our 
main result, solution FA02, utilizes a complete set of 
forward and fixed-t dispersion relation constraints, 
from threshold to 1~GeV, and from t = 0 to $-0.4~(GeV/c)^2$, 
applied to the $\pi$N elastic amplitude.  A large number 
of systematic checks have been performed, including 
fits with no charge-exchange data and other database 
changes, fits with few or no dispersion relation 
constraints, and changes to the Coulomb correction 
scheme.  We have also reexamined methods used to 
extract Breit-Wigner resonance parameters. The quality 
of fit to both data and dispersion relation constraints 
is superior to our earlier work.  The results of these 
analyses are compared with previous solutions in terms 
of their resonance spectra and preferred values for 
couplings and low-energy parameters, including the $\pi NN$
coupling constant. 
\end{abstract}

\pacs{14.20.Gk, 13.30.Eg, 13.75.Gx, 11.80.Et}

\maketitle

\section{Introduction}
\label{sec:intro}

The determination of resonance properties for all accessible
baryon states is a central objective in Nuclear Physics. 
Pole positions, decay channels, and branching ratios have
been extracted from hadronic reaction data. Further results 
for magnetic moments and photo-decay amplitudes have been 
obtained from real and virtual photon interactions.  This 
body of information has been used to test QCD-inspired 
models and, more recently, lattice calculations.

Most $N$ and $\Delta$ resonances, listed as 3- and 4-star
states in the Review of Particle Properties (RPP)~\cite{rpp}, 
have had their existence, masses, and widths determined
through single-channel fits to scattering data, with
$\pi$N elastic scattering being the predominant source. 
Similarly, most photo-decay amplitudes have been determined 
in fits to single-pion photoproduction.  However, a number 
of recent studies of eta photoproduction and 
electroproduction data have claimed~\cite{eta_photo}
resonance properties for the N(1535) and N(1520) at 
variance with these long-standing pion production results. 
Multi-channel analyses have provided further estimates~
\cite{multi} which also tend to disagree with the 
single-channel values. 

These problems have motivated an improved analysis of $\pi$N 
elastic scattering in the $\eta$N threshold region.  Here, we 
report the results of a coupled-channel fit to $\pi$N elastic 
and $\eta$N production data, the $\pi N\to\eta N$ data having 
been incorporated through a Chew-Mandelstam K-matrix 
formalism~\cite{fa84}.  Compared to our previous work 
(solution SM95~\cite{sm95}), the present FA02 analysis 
has fitted a larger database, in particular new 
high-quality data in the $\Delta$(1232) resonance 
region (described in Section~\ref{sec:expt}).
In addition to the forward C$^\pm (\omega)$ and fixed-t
B$^\pm (\nu , t)$ dispersion relation constraints
already used in SM95, our new solution has been 
constrained by further forward derivative E$^\pm 
(\omega)$ and fixed-t C$^\pm (\nu , t)$ dispersion 
relations~\cite{hoehler} over broader ranges of 
energy and four-momentum transfer.

In Section~\ref{sec:form}, we will explain how the 
dispersion relation constraints were implemented.  For 
comparison purposes, fits to the available data were 
also performed with fewer, and no, dispersion relation 
constraints.  This has helped to gauge the relative 
effect of such constraints on the quality of fit to 
data. Many other fits were performed to study systematic 
effects on resonance and dispersion relation parameters, 
in particular the $\pi NN$ coupling constant. We have 
performed fits excluding charge-exchange data, employing 
different datasets in the $\Delta(1232)$ region, using 
different Coulomb correction schemes, and retaining the 
older Karlsruhe~\cite{koch} D-waves at low energies, to 
gauge the influence of these poorly known amplitudes.

In Section~\ref{sec:sm95}, we compare the present FA02 
solution with both our previous SM95 solution and 
results from previous analyses, as well as giving values 
for the partial-wave amplitudes. The baryon spectrum, 
and associated couplings are given in Section~
\ref{sec:res}, whereas results for the $\pi$NN coupling 
constant and other dispersion relation parameters are 
discussed in Section~\ref{sec:drResults}. In Section~
\ref{sec:conc}, we provide a brief summary and consider 
what extensions of this work can be expected in the future.

\section{Database}
\label{sec:expt}

\begin{figure}[th]
\centering{
\includegraphics[height=0.7\textwidth, angle=90]{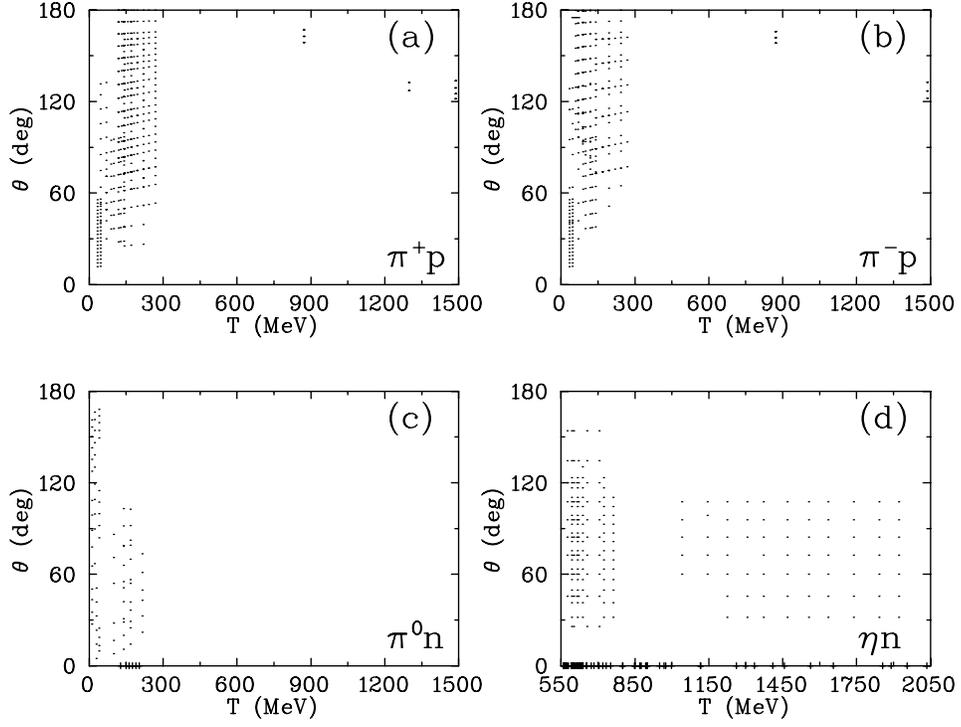}
}\caption{Energy-angle distribution of recent (post-1995)
      data: (a) $\pi^+p$ and (b) $\pi^-p$ elastic   
      scattering, (c) charge-exchange $\pi^-p\to\pi^0n$,
      and $\pi^-p\to\eta n$ (all available).  Total
      cross sections are plotted at zero degrees.
      \label{fig:g1}}
\end{figure}

The evolution of our database is summarized in Table~
\ref{tbl1}.  Over the course of four previous 
pion-nucleon analyses~\cite{fa84,sm90,fa93,sm95} our 
energy range has been extended from 1.1 to 2.1~GeV. 
Additions to the current database~\cite{said} are due 
mainly to measurements in the low (below 300~MeV) 
energy region.  We have also incorporated 173 $\eta$ 
production data (from the process $\pi^-p\to\eta n$).  
Below, we list the recent (post-1995) additions for 
elastic and charge-exchange scattering.  The 
eta-production database, being recently added, contains 
both new and old measurements.  Here, we should note 
that the full database contains conflicting results.  
Some of these with very large $\chi^2$ have been 
excluded from our fits.  However, all available data 
have been retained (the excluded data labeled as 
``flagged"~\cite{flag}) so that comparisons can be made 
through our on-line facility~\cite{said}.  Some individual 
data points were also removed from the analysis in order 
to resolve conflicts or upon authors' requests (since our 
previous analysis~\cite{sm95}, we have flagged 48 $\pi^+$ 
and 35 $\pi^-$ data, most of these being total cross 
sections).  Some of the data, listed as new, were 
available in unpublished form at the time of our 
previous analysis~\cite{sm95}.  A complete description 
of the database and those data not included in our fits 
is available from the authors~\cite{said}.  The 
energy-angle distribution of recent (post-1995) $\pi^+p$, 
$\pi^-p$ elastic, $\pi^-p\to\pi^0n$, and all 
$\pi^-p\to\eta n$ data is given in Fig.~\ref{fig:g1}.

Since most of the new data~
\cite{pa01,br95,ho98,pa02,ho03,fr96,ga99,is99,kr99,fr98,ja97,ra96,jo95,wi96,al95,al00}
are from high-intensity facilities (such as LAMPF, 
TRIUMF, and PSI), they generally have smaller 
statistical errors and, thus, have a greater influence 
on the fits.  As mentioned above, a large fraction of 
the more recent $\pi^\pm$p data were produced at 
energies spanning the $\Delta(1232)$ resonance.  These data 
have been taken mainly at TRIUMF.  From this source, we have 
added 106 $\pi^+p$ and 54 $\pi^-p$ differential cross 
sections, from 140 to 270~MeV at medium scattering 
angles~\cite{pa01}, and 41 $\pi^+p$ (37 $\pi^-p$) 
cross sections, from 90 to 140~MeV~\cite{br95}.  A 
further 194 $\pi^+p$ and 81 $\pi^-p$ $A_y$ data between 
90 and 280~MeV, at medium and backward scattering 
angles, have been collected using the CHAOS facility 
at TRIUMF~\cite{ho98,ho03}.  At low energies, 55 to 140~MeV, 
and backward scattering angles, 13 $\pi^+p$ and 89 $\pi^-p$ 
$A_y$ data have also become available from CHAOS~\cite{pa02}. 

A few LAMPF experiments have now been analyzed and added to 
our database.  These include 36 high-quality polarized 
charge-exchange data between 100 and 210~MeV~\cite{ga99}, 40 
low-energy (10 to 40~MeV) cross sections~\cite{is99}, and 6 
forward charge-exchange differential cross sections at 
27~MeV~\cite{fr98}.

We have added 23 $\pi^+p$ and 6 $\pi^-p$ TRIUMF~\cite{fr96}
and 44 $\pi^+p$ and 15 $\pi^-p$ LAMPF~\cite{kr99} partial 
total cross sections from 40 to 280~MeV~\cite{fr96}, though 
this quantity is not fitted in the analysis.

PSI experiments have provided 5 $\pi^-p$ differential cross 
sections in the backward hemisphere between 45 and 70~MeV~
\cite{ja97} and 11 $A_y$ data at 160 and 240~MeV~\cite{ra96}.  
After a revised analysis, the Karlsruhe group has published 
a final set of both $\pi^\pm$p low-energy differential 
cross sections~\cite{jo95} and analyzing powers~\cite{wi96}.  

A set of polarization parameters $P$, $R$, and $A$ for both 
$\pi^+p$ and $\pi^-p$ from 870 to 1490~MeV were contributed 
by the ITEP--PNPI Collaboration~\cite{al95,al00}.  

In the $\Delta(1232)$ resonance region, there are two sets 
of data which are in disagreement, primarily on the rising 
side of the resonance: the first includes the PSI total 
cross section data of Pedroni \textit{et al.}~\cite{ped78} 
and the differential cross sections of Brack \textit{et al.}~
\cite{br86,br95}, while a second set includes the total 
cross sections of Carter \textit{et al.}~\cite{ca71} and
the differential cross sections of Bussey \textit{et al.}~
\cite{bu73}.  Since the resonant $\Delta(1232)$ (P$_{33}$) 
amplitude dominates most dispersion relations, discrepant 
data bases in this energy region are no small concern.  In 
our previous solution~\cite{sm95}, the Bussey \textit{et al.}
data were assigned 5\% normalization uncertainties, and the 
Carter \textit{et al.} data were assigned 1.5\% uncertainties, 
to resolve a discrepancy between the energy-dependent and 
single-energy solutions. Subsequent to that solution, the 
normalization uncertainties in the Bussey \textit{et al.} 
and Carter \textit{et al.} data were reevaluated (to 1\%), 
and the lowest energy measurements in each of the $\pi^+$p 
and $\pi^-$p total cross sections were removed, by one of 
the principal authors~\cite{bugg-renorm}. The data were so 
employed in the present solution.  However, with the addition 
of new TRIUMF differential cross section~\cite{pa01} and 
polarization data~\cite{ho98,ho03}, the changes in phases and 
dispersion relation parameters arising from selecting one 
of the aforementioned discrepant data sets have greatly
diminished, as will be discussed in Sec.~\ref{sec:drResults}. 
All of the above data sets are included in the 
fitted database.

Most measurements of the $\pi^-p\to\eta n$ reaction cross 
section are rather old and sometimes conflicting.  There 
are few cross section measurements above 800~MeV and no 
polarized measurements below 1040~MeV~\cite{said}.  We 
have selected 138 differential and 35 total cross 
sections for our analysis, with an additional 176 data 
points added to the on-line database but not included 
in this study~ 
\cite{e909,bi73,bu69,de75,de69,ri70,br79,fe75,pe64,ba71,wa68,ka75,ne72}.  
A detailed analysis of the older data can be found in the 
review by Clajus and Nefkens~\cite{clnef}.  Previous 
unpolarized measurements are listed in Ref.~\cite{list}.  
One new set~\cite{e909} of data has been included in our 
analysis.  Total and differential cross sections for 
$\pi^-p\to\eta n$, near threshold, were measured using the 
BNL-AGS Facility~\cite{e909}.  Cross sections were obtained 
from threshold (560~MeV) to $\sim$650~MeV with $\sim$10\% 
statistics for the total cross sections.  The angular 
coverage was 45$-$155$^{\circ}$.  Systematic uncertainties 
were estimated at $\sim$3\%, a value significantly less 
than claimed in previous measurements, where systematics 
were above 10\%.

\section{Formalism}
\label{sec:form}

There are three principal components to the methodology we 
use in fitting $\pi N$ elastic scattering, charge exchange 
and $\pi^-p\to\eta n$ data: (i) the energy-dependent 
parametrization of partial waves, (ii) dispersion-relation 
constraints, and (iii) Coulomb interaction effects.

\subsection{Chew-Mandelstam K-matrix}
\label{sec:chma}

Our energy-dependent partial-wave fits are parameterized 
in terms of a coupled-channel Chew-Mandelstam K-matrix, as 
described in Refs.~\cite{sm95,fa84}.  This choice 
determines the way we introduce and modify the 
energy-dependence and account for unitarity in our fits.  

The scattering into different channels is represented by a
matrix $\bar T$, parameterized in terms of $\bar K$, a
(4$\times$4) real symmetric K-matrix for each partial wave  
\begin{eqnarray}
\bar T = \bar K (1 - C \bar K )^{-1},
\label{1}\end{eqnarray}
$C$ being a diagonal Chew-Mandelstam function, with ${\rm 
Im} C_i = \rho_i$ giving the phase-space function for the 
i$^{\rm th}$ channel ($\pi N$, $\eta$N, $\pi\Delta$, and 
$\rho$N).  The T-matrix for elastic $\pi N$ scattering is 
then given by $T_{\pi\pi} = \bar T_{\pi\pi} \rho_{\pi}$; 
for $\pi N\to\eta N$, the relation being $T_{\pi\eta} = 
\bar T_{\pi\eta} \left(\rho_{\pi} \rho_{\eta}\right)^{1/2}$. 
The $\rho$N channel is new to the present solution. Its
inclusion has reduced the number of parameters by 6, compared to
SM95~\cite{sm95}, while improving the fits and dispersion relation
consistency (the $\rho$ width is accounted for in this
approach).  It should
be noted that the above $\pi\Delta$ and $\rho N$ channels
are generic and included to preserve unitarity.  Unlike
the SM95 fit, however, the $\eta$N channel has been fitted
to physical cross sections.

In order to control the behavior of each T-matrix near 
threshold, the K-matrix elements were expanded as 
polynomials in the energy variable $Z = (W - W_{th})$, 
where $W$ and W$_{th}$ are the center-of-mass ($\sqrt{s}$) 
and threshold energies, respectively.  Multiplying by an 
added factor of $Z$ allowed the fixing of scattering 
lengths through the value of the leading term.  

Single-energy analyses were parametrized as
\begin{eqnarray}
S_e = (1 + 2iT_e) = \cos (\rho) ~ e^{2 i \delta},
\label{2}\end{eqnarray}
with the phase parameters $\delta$ and $\rho$ expanded as 
linear functions around the analysis energy, and with a 
slope (energy derivative) fixed by the energy-dependent 
solution.  It should be emphasized that the single-energy 
solutions are generated mainly to search for missing 
structure.  The scatter around an energy-dependent 
solution is also useful as a measure of the uncertainty in 
a partial wave.  The variation from point-to-point should 
not be taken as proof of sharp structures, as these binned 
fits are far less constrained than the global fits (they do 
not, for example, satisfy the dispersion relation 
constraints)~\cite{gerhard}.

Threshold behavior was determined in the following manner.  
The S-wave scattering lengths were linked to our dispersion 
relation constraints, as described below.  The P-wave 
scattering volumes were searched without constraint, but it 
was ensured that near threshold these waves followed the 
appropriate Chew-Low forms~\cite{chew-low,hamilton}, which
are approximations to the respective partial-wave dispersion
relations. The D-waves were weakly constrained to the older 
Karlsruhe analysis by assigning  
5\% errors to the Koch values~\cite{koch}, and the low-energy 
behavior of higher waves was fixed to Koch's result~
\cite{koch}.  Once an appropriate hadronic amplitude was 
determined, charge corrections were applied as described in 
Sec.~\ref{sec:clmb}.  

\subsection{Dispersion Relation Constraints}
\label{sec:disp}

\begin{figure}[th]
\centering{
\includegraphics[height=0.7\textwidth, angle=90]{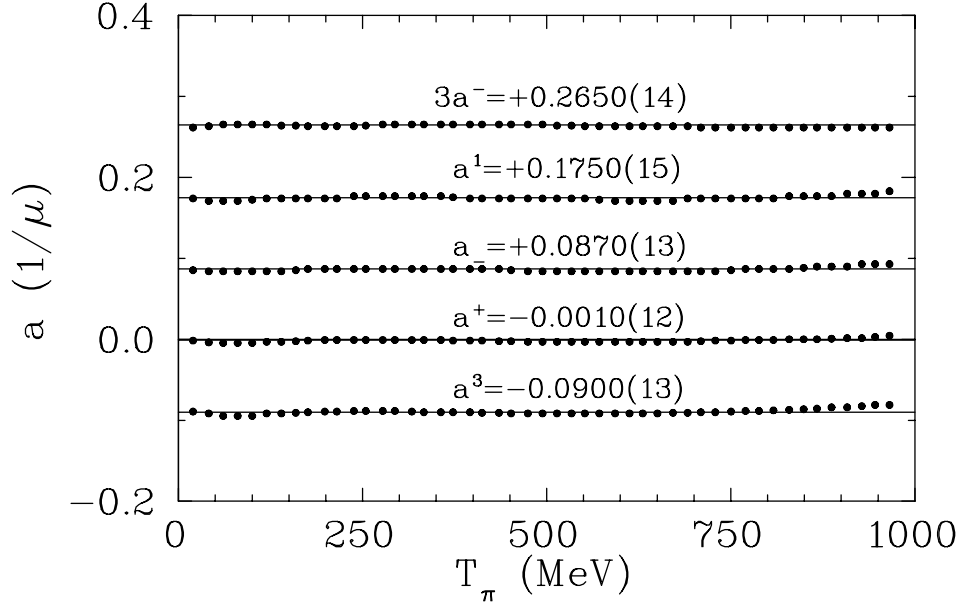}
}\caption{Subtraction constants as a function of pion   
          kinetic energy for the forward C$^\pm$ dispersion
          relations plotted in terms of the S-wave
          scattering lengths $a^\pm$, and other combinations,
          in inverse pion mass units.  Horizontal
          lines represent the least-square averages of
          individual values. \label{fig:g2}}
\end{figure}

In general, a fit of the K-matrix elements, 
expanded in terms of an energy variable, may not result in a 
form satisfying all of the requirements imposed by analyticity 
and crossing symmetry. In our analysis, those requirements are 
addressed at fixed four-momentum transfer $t$ by a complete 
set of fixed-t dispersion relations (DRs), which are handled 
iteratively with the data fitting, as has been described in
Ref.~\cite{sm95}. The DRs contain subtraction 
constants which should be independent of energy (but which can 
be functions of four-momentum transfer).  After each 
data-fitting iteration, these constants are calculated as a 
function of energy.  $\chi^2$ deviations from the average, at 
a series of energies, are then calculated and included as 
pseudo-data.  Figure~\ref{fig:g2} shows an example of 
the subtraction constants in the forward C$^\pm (\omega 
)$ DRs up to 1~GeV, $\omega$ being the pion lab energy.
The partial-wave amplitudes and the real parts 
of the DR invariant amplitudes are then adjusted to minimize 
the $\chi^2$ from the sum of data and pseudo-data.  
Compatibility with the DR constraints can be controlled 
through the errors assigned to pseudo-data in the fit.

In our previous published analysis~\cite{sm95}, we employed 
constraints from fixed--t DRs for the 
B$_\pm (\nu ,t)$, $\nu$ equal to $(s-u)/4M$,
and forward C$^\pm (\omega )$, invariant 
amplitudes~\cite{notation}.
Constraints on the partial-wave fits were generated at fixed 
values of $t$ ranging from 0 to $-0.3~(GeV/c)^2$ and pion 
lab kinetic energies from 25 to 600~MeV.  The B$_\pm (\nu 
,t)$ DRs were constructed in the H\"uper form~\cite{hoehler}. 
The full $\pi$N amplitude was parameterized by two invariant 
amplitudes B$^\pm (\nu ,t)$ and C$^\pm (\nu ,t)$ (or 
equivalently B$^\pm (\nu ,t)$ and A$^\pm (\nu ,t)$). 
While in the analysis of Ref.~\cite{sm95} the C$^\pm$ 
amplitudes were reasonably well behaved, improvement called 
for their inclusion in DR constraints at non-forward angles. 

With the inclusion of constraints from the fixed--t C$^\pm 
(\nu ,t)$ DRs, we complete the fixed-t DRs coverage of the 
full $\pi$N amplitude. The C$^\pm$ DRs were chosen since 
parameterizations of the very high energy amplitudes are 
readily available~\cite{hoehler}.  The energy and 
four-momentum transfer coverage was increased for most 
fixed-t DRs to $T_{\pi}< 1000$~MeV and $0\le -t\le 0.4~
(GeV/c)^2$.  This $t$ range covers the entire angular range 
up to $\sim$450~MeV, across the important $\Delta$(1232) 
resonance region, which contains the most accurate and 
complete data sets.  Due to use of analytically smooth 
energy-dependent amplitudes, the DRs are approximately 
satisfied over wider constraint energy and $t$ ranges
(see results in Section~\ref{sec:drResults}).

\begin{figure}[th]
\centering{
\includegraphics[height=0.4\textwidth, angle=90]{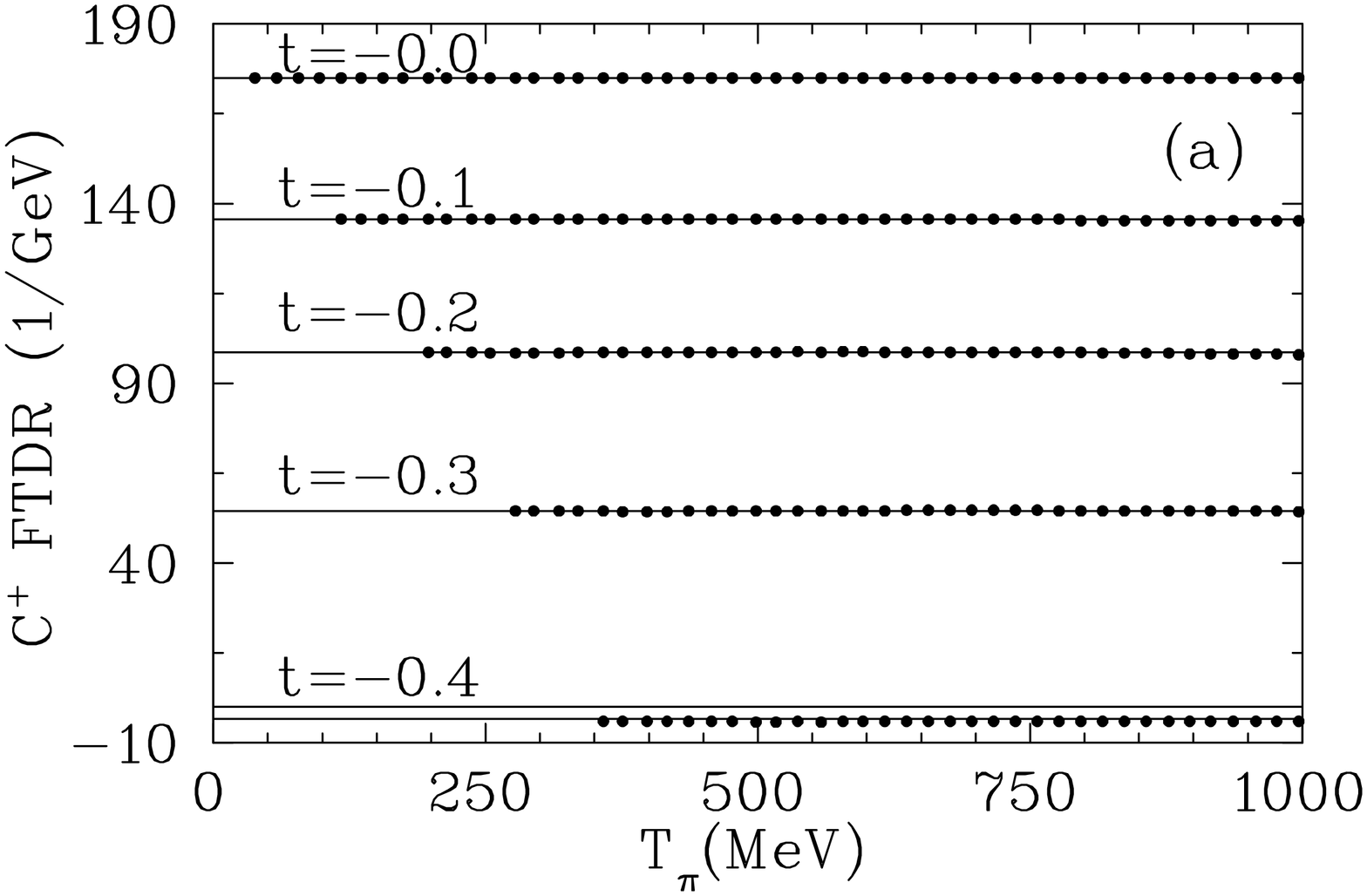}\hfill
\includegraphics[height=0.4\textwidth, angle=90]{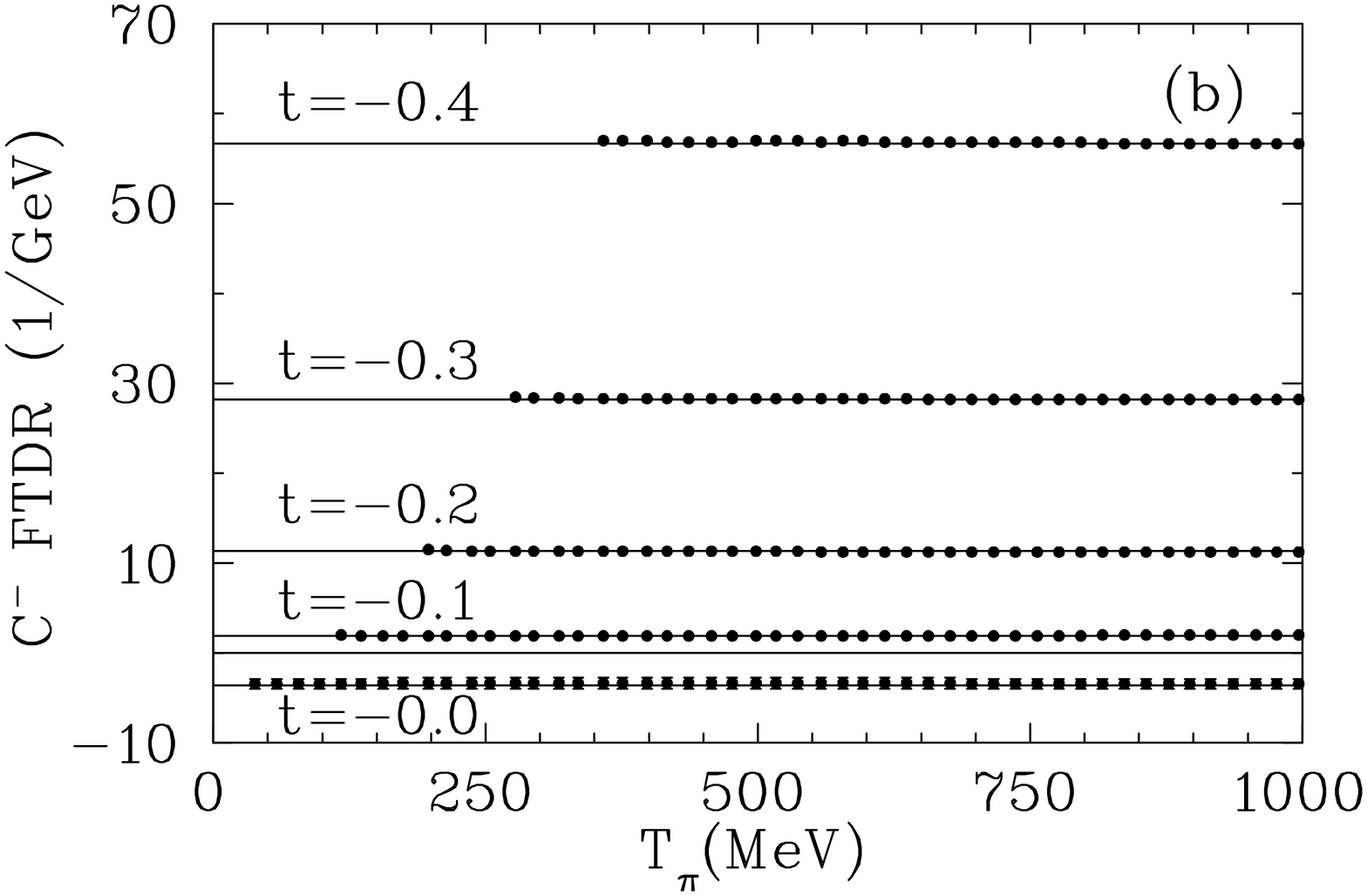}
}\caption{Subtraction constants as a function of pion
          kinetic energy at four values of four-momentum
          transfer $t$ [in $(GeV/c)^2$] for the fixed-t
          (a) C$^+$($\nu$, t) and (b) C$^-$($\nu$, t)
          dispersion relations.
          Horizontal lines represent the least-square   
          averages of individual values. \label{fig:g3}}
\end{figure}

In addition to the fixed--t C$^\pm (\nu ,t)$ DRs, the 
present analysis includes constraints from the forward 
derivative DRs 
\begin{eqnarray}
E^\pm (\omega) = \frac{\partial}{\partial t} 
[A^\pm (\nu,t) + \omega B^\pm (\nu,t) ] |_{t=0} 
\label{3}\end{eqnarray}
utilized in Ref.~\cite{glls88}.  These DRs were also applied 
from $10 < T_{\pi} < 1000$~MeV.  The E$^\pm (\omega)$ DRs 
provide a stronger constraint on higher partial waves, 
which are more poorly known at low energies.  Nevertheless, 
the $l = 1$ $\Delta$(1232) resonance gives
a dominant 
contribution, as is the case for most of the other DRs we
employ.  Consistency 
of the fixed--t C$^\pm (\nu, t)$ DRs is illustrated in 
Fig.~\ref{fig:g3}.  As in Fig.~\ref{fig:g2}, the plotted 
constant is obtained through a delicate cancellation 
between pole contributions and principal value integrals. 
The variation of these contributions is displayed for the 
forward E$^\pm (\omega)$ DRs in Fig.~\ref{fig:g4}.

Dispersion relations contain principal value integrals over 
the imaginary parts of invariant amplitudes up to infinite 
energy.  The subtraction constants are used to ensure 
convergence of these integrals.  As our analysis extends up 
to 2100~MeV, the amplitudes above this energy are obtained 
from other sources.  For the energy region 2100$-$5000~MeV, 
the Karlsruhe KA84 $\pi$N analysis solution~\cite{ka84} is 
utilized.  At higher energies, parameterizations were 
taken from Ref.~\cite{hoehler}.  These high-energy 
contributions are relatively small for the H\"uper DR and 
isovector amplitudes, but are important for the isoscalar 
E$^+$ and C$^+$ amplitudes.  In these latter cases, while 
high-energy contributions were necessary to achieve 
sufficient convergence, the various forms of 
parameterization~\cite{hoehler} yielded negligible 
differences in the DR constraints.

As mentioned above, DRs contain a number 
of parameters which are \textit{a priori} unknown, namely 
the various subtraction constants (if inserted for 
convergence of the DR integrals) and the $\pi NN$ coupling 
constant $g^{2}/4\pi$ (when the Born term appears).  These 
constants are considered free parameters to be determined 
from the fits to the data and the DR 
constraints.  In principle, these subtraction constants can 
be determined at any energy.  We fix them at the physical 
threshold for the forward DRs, and at the unphysical points 
C$^\pm (0, t)$ for the fixed-t C$^\pm$ DRs.  The former can 
be expressed in terms of the S-wave scattering lengths 
a$^\pm$ (C$^\pm$ DRs) and the P-wave scattering volumes 
a$_{1+}^\pm$ (E$^\pm$ DRs).

These parameters could be determined in a number of ways.  In 
the present study, we settled on the following procedure: for 
a particular DR and data minimization run, the $\pi^-p$ S-wave 
scattering length $a_-$ was fixed around  
precise results extracted from the PSI 
pionic-atom experiment~\cite{piat}, and the coupling 
constant $g^2/4\pi$ and scattering volume $a_{1+}^+$ were 
fixed at chosen values.
In essence, the $\pi^- p$ S-wave scattering length $a_-$ is
used as an additional data point at threshold.
The other parameters (including the 
isovector S-wave scattering length $a^-$) were allowed to 
float freely.  As a result, the subtraction constant was fixed 
for the forward C$^+$ and E$^+$ DRs, with real parts of the 
amplitudes forced to particular values in the analysis, 
whereas analogous constants for the other DRs were allowed to 
find their own level. Once a solution with minimal $\chi^2$ 
was determined, further runs were performed with new sets of 
fixed ($g^2/4\pi$, $a_{1+}^+$) values, producing a mapping 
over a grid of parameters.  The solution FA02 was selected 
from this grid of solutions based on overall $\chi^2$.  

Generally, the $\chi^2$ variation of parameters is parabolic 
in the region near a minimum.  The advantage of a grid method 
is that the depth of the ($g^2/4\pi$, $a_{1+}^+$) minimum in 
$\chi^2$ space indicates how well these parameters are 
determined.  One can also view the dependence of the $\chi^2$ 
in each $\pi$N charge channel, or the DR pseudo-data, 
separately to look for systematic variations.   

We found the mapping procedure to be
necessary for ($g^2/4\pi$, $a_{1+}^+$) 
due to computational instabilities which arise occasionally.  One 
could, in principle, fix all the parameters and follow the above 
procedure over a multi-dimensional grid, and tests were done to 
investigate this, but it was found that the above procedure was 
sufficient to ensure stable fits with well-satisfied DRs in all 
cases, while keeping the $\chi^2$ fit to the data optimal.  
Adding more, or fixing more constraints, increases the data fit
$\chi^2$. 

\subsection{Coulomb Interaction}
\label{sec:clmb}

The partial-wave and DR analyses are performed 
using purely isospin-conserving hadronic amplitudes, so Coulomb and 
other isospin-violating contributions must be introduced  before 
constructing observables to compare to the physical pion-proton 
scattering data.  There is no unique way to split the interaction 
into strong and electromagnetic parts~\cite{gass03}, so in general 
models must be used.

In previous analyses, we have used a simple point-source barrier 
factor correction for all partial waves at all energies, as 
discussed in Ref.~\cite{fa84}.  The main isospin-violating 
contribution is the $\Delta^{++}-\Delta^0$ mass and width difference, 
which we have been treating phenomenologically by splitting the 
average resonant hadronic $P_{33}$ partial-wave into $P^+_{33}$ and 
$P^-_{33}$ waves for $\pi^+p$ and $\pi^-p$, respectively. The $\Delta$ 
mass and width differences were found from a best fit to the data.  A 
further correction was made to the $P_{33}$ amplitude to account for 
the $\pi^-p\to\gamma n$ channel. The mass ($\Delta^0 -\Delta^{++}$)
and width ($\Gamma^0 - \Gamma^{++}$) differences are 1.74$\pm$0.15~MeV
and 1.09$\pm$0.64~MeV, respectively~\cite{bugg}.

The main criticism of point-source barrier factors is that they 
exaggerate the S-wave corrections at low energies, since the $l=0$ 
partial waves do not vanish at the origin, while the point-Coulomb 
field diverges there.  Consequently, in this study, 
we have used the Nordita~
\cite{trom} Coulomb prescription, which was used for the 
Karlsruhe-Helsinki (KH80) analyses~\cite{kh80}.  However, the 
Nordita work supplied corrections for only the largest S- and P-waves 
up to 530~MeV, whereas our analysis extends to 2100~MeV. The Nordita 
corrections were therefore supplemented with corrections calculated 
from Gibbs extended-source barrier factors~\cite{gibbsCoul} for S- 
and P- waves at energies above 450~MeV, and D- and higher waves at all 
energies. The Nordita and Gibbs results were smoothly joined in the
450~MeV region.
As well, an error was found and corrected in the isospin 
mixing term in one Nordita source~\cite{trom}, while the  pure Coulomb 
part of the (dominant) resonant $P_{33}$ correction was reduced 15\% 
to take into account intermediate $\sigma$ exchange~\cite{buggCoul}, 
which was missing in the Nordita work. 

The Nordita corrections already contain effects due to the $\pi^+
-\pi^0$ mass difference. In this scheme, the $\pi^+$ channel 
corresponds to the hadronic channel, and all isospin violating effects 
(due to mass differences and the $\gamma n$ channel) are subsumed in 
the $\pi^-$ amplitudes.  To remain consistent with this scheme, we 
have redefined our $\Delta$(1232) resonance charge-splitting 
analogously.  This has had the effect of shifting the peak of the 
hadronic resonance to a slightly lower energy.

\begin{figure}[th]
\centering{
\includegraphics[height=0.4\textwidth, angle=90]{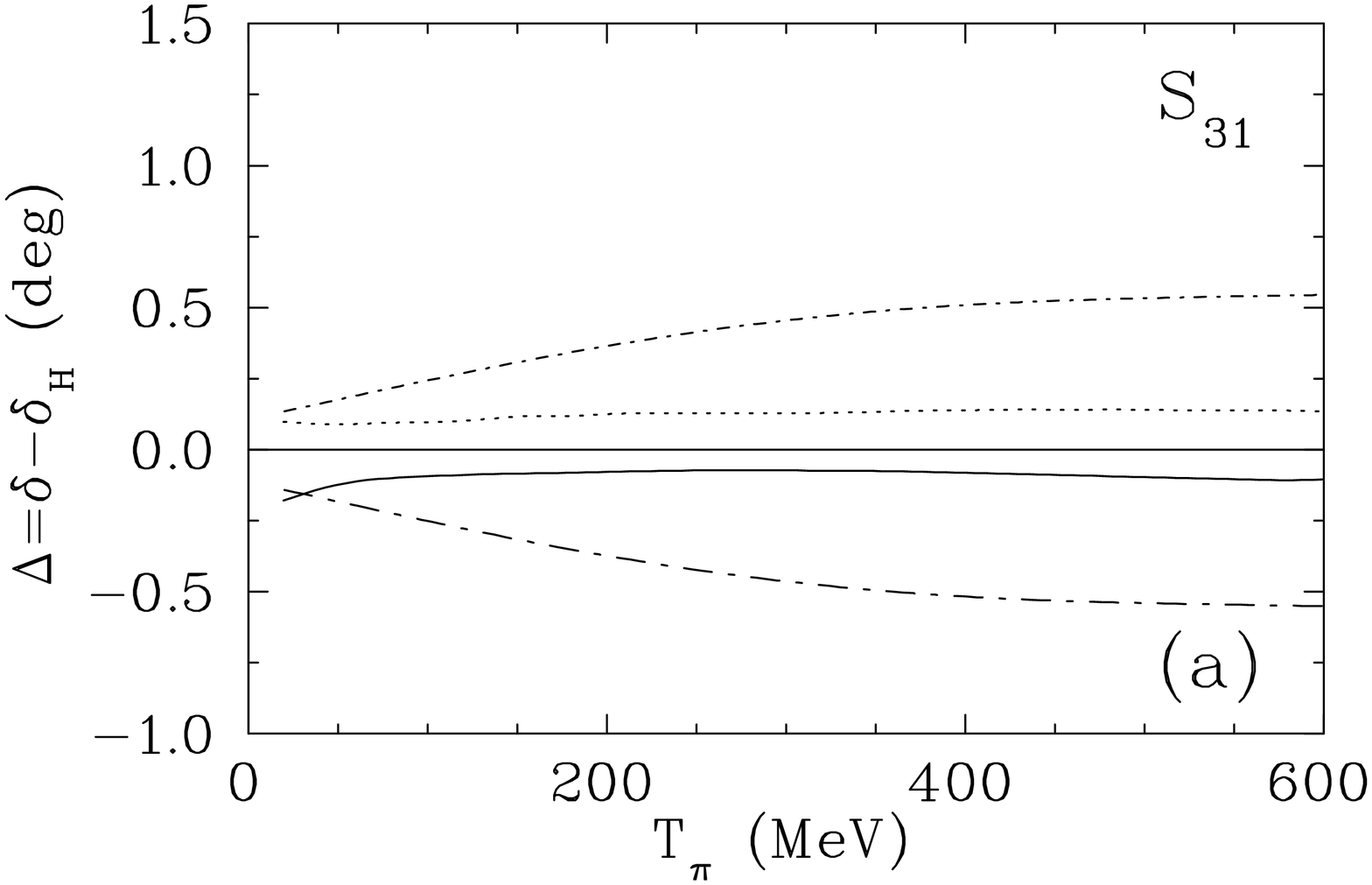}\hfill
\includegraphics[height=0.4\textwidth, angle=90]{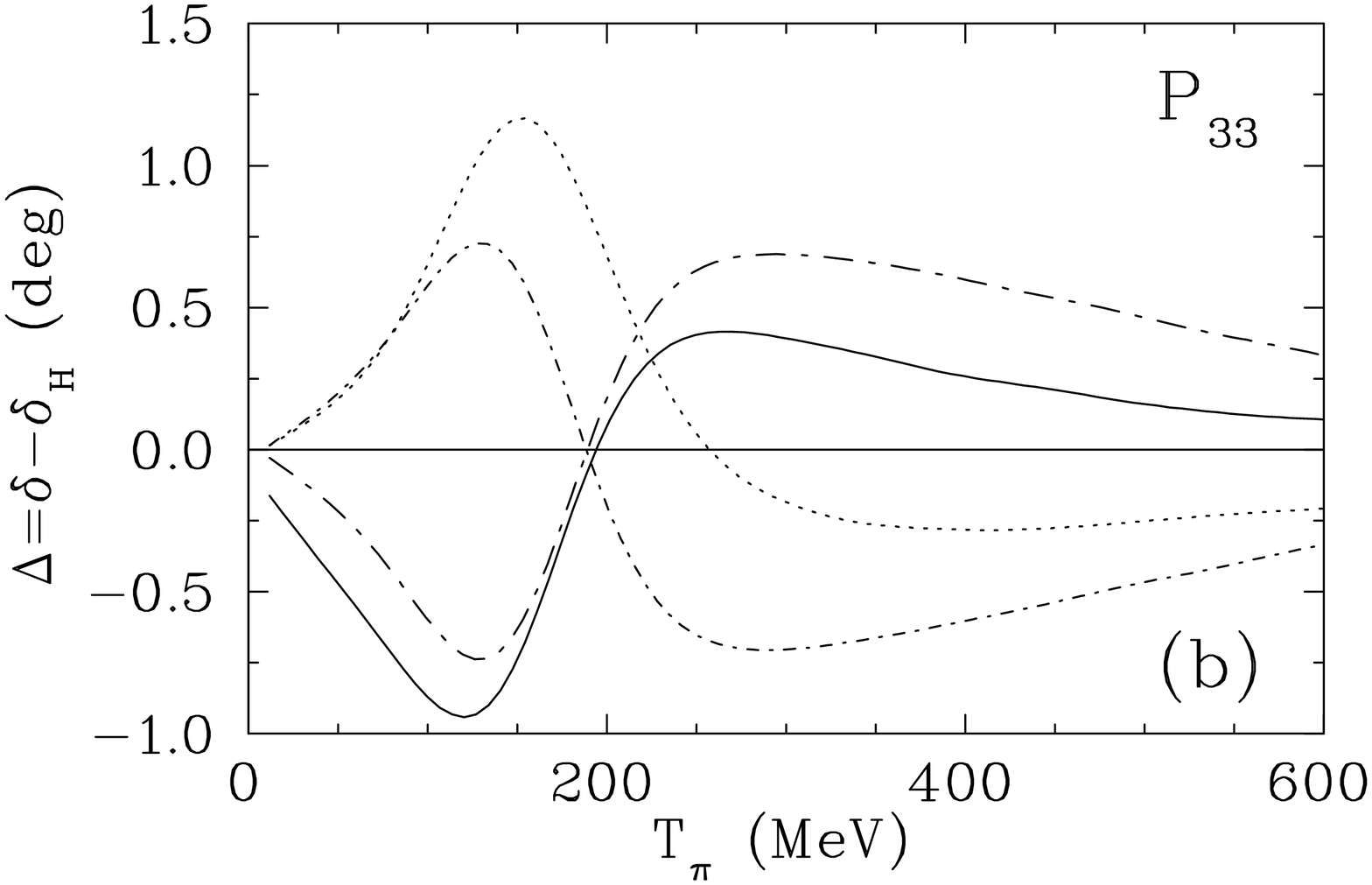}
}\caption{Comparison of the point-Coulomb (SM95) and 
          Nordita-Gibbs (FA02) charge corrections $\Delta 
          = \delta - \delta_H$, the subscript $H$ denoting 
          an hadronic phase, for the (a) $S_{31}$ and (b) 
          $P_{33}$ partial waves.  In both plots, 
          dash-dotted and short-dash-dotted (solid and 
          dotted) curves give point-Coulomb~\protect\cite{sm95} 
          (FA02) corrections for $\pi^+p$ and $\pi^-p$, 
          respectively. \label{fig:g5}}
\end{figure}

Final FA02 Coulomb correction results for the S$_{31}$ and P$_{33}$ 
partial-wave are plotted in Fig.~\ref{fig:g5}, along with the 
point-source barrier factor results.  Differences between the FA02 
and point-source corrections (not shown) are similarly large in the 
S$_{11}$ partial-wave, as expected, while for all other waves, the 
differences are much smaller.  However, these differences in the 
SM95 and FA02 correction schemes did not result in large effects
on the fits or dispersion relation parameters.

\section{Results and Discussion}
\label{sec:output}

\subsection{FA02 versus the SM95 Fit}
\label{sec:sm95}

The main result of this work has been an energy-dependent solution 
(FA02) having a $\chi^2$ of 45874 for 23979 ($\pi^\pm$ elastic, with 
pion induced $\pi^0$ and $\eta$ production) data to 2.1~GeV ($\eta$N 
data have been included to 800~MeV).  The overall $\chi^2$/data was 
significantly lower than that found in our previously published fit 
($\chi^2$/data = 2.2)~\cite{sm95}, despite the inclusion of 
additional data.  This change is partly a 
reflection of the database changes mentioned in Section~
\ref{sec:expt}.  Our present and previous energy-dependent solutions 
are compared in Table~\ref{tbl1}.  As in previous analyses, we have
used the systematic uncertainty as an overall normalization factor 
for angular distributions.  For total cross sections and excitation 
data, we have combined statistical and systematic uncertainties in 
quadrature.  This renormalization freedom allows a significant 
improvement in our best-fit results, as shown in Table~\ref{tbl2}.  
In cases where the systematic uncertainty varies with angle, this 
procedure may be considered a first approximation.

In Table~\ref{tbl3}, we compare the energy-dependent and single-energy 
results over the energy bins used in these single-energy 
analyses.  Also listed is the number of parameters varied in each 
single-energy solution.  A total of 172 (94 for isospin $1/2$ and 
78 for isospin $3/2$) parameters were varied in the energy-dependent 
analysis FA02.  The extended database has allowed an increase in the 
number of single-energy values 
versus our previous result~\cite{sm95} over the same 
energy range.

\begin{figure}[th]
\centering{
\includegraphics[height=0.4\textwidth, angle=90]{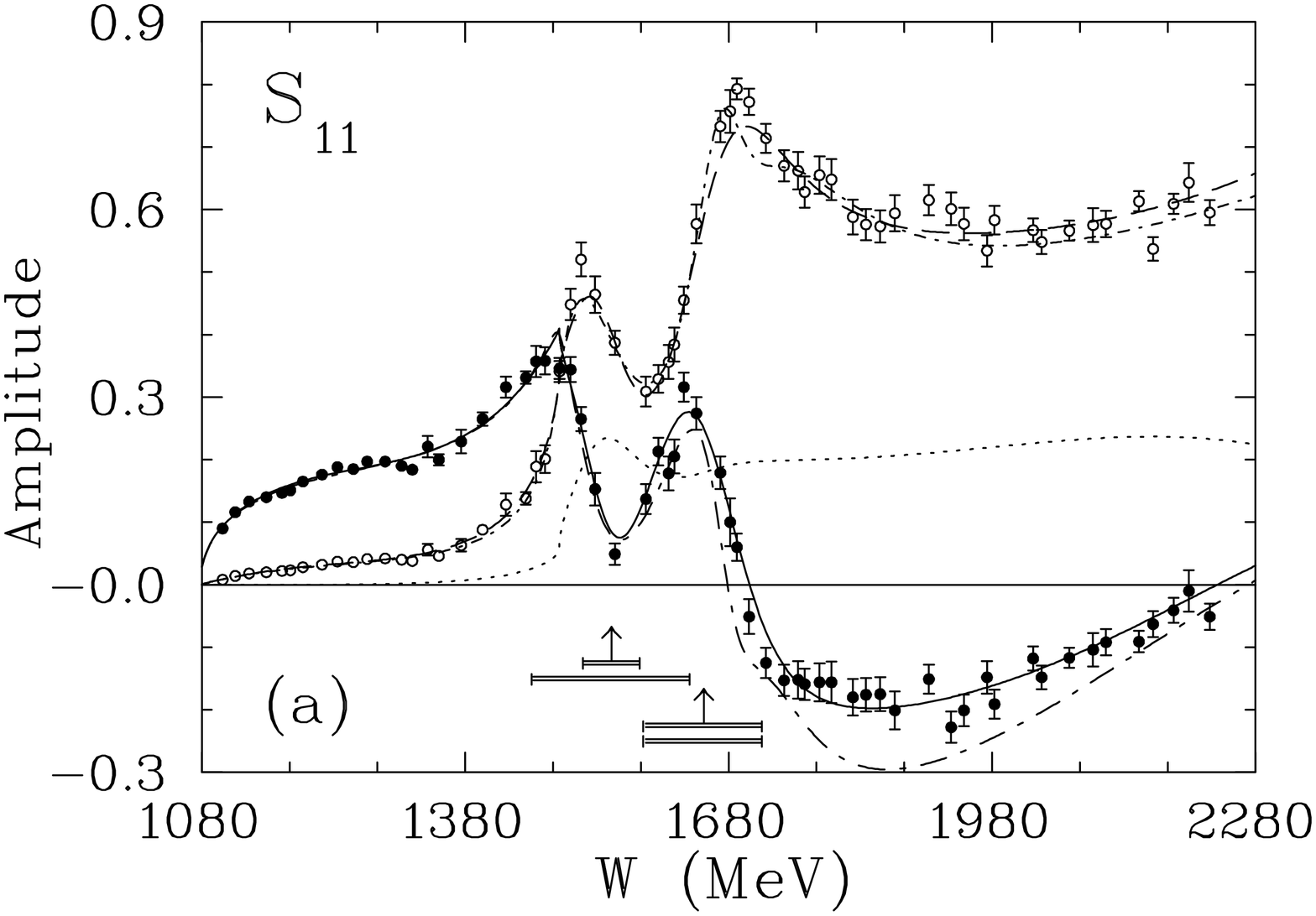}\hfill
\includegraphics[height=0.4\textwidth, angle=90]{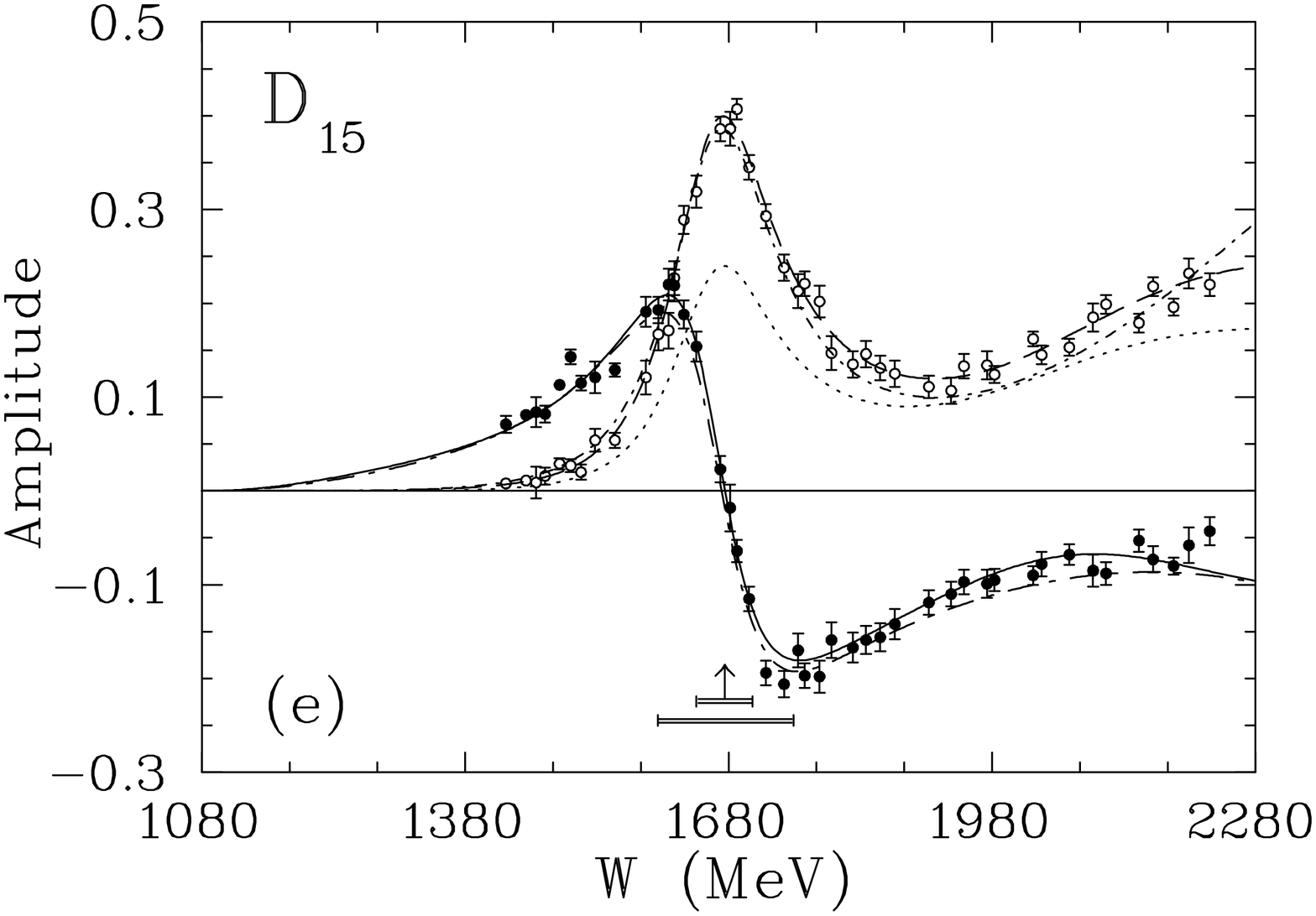}
\includegraphics[height=0.4\textwidth, angle=90]{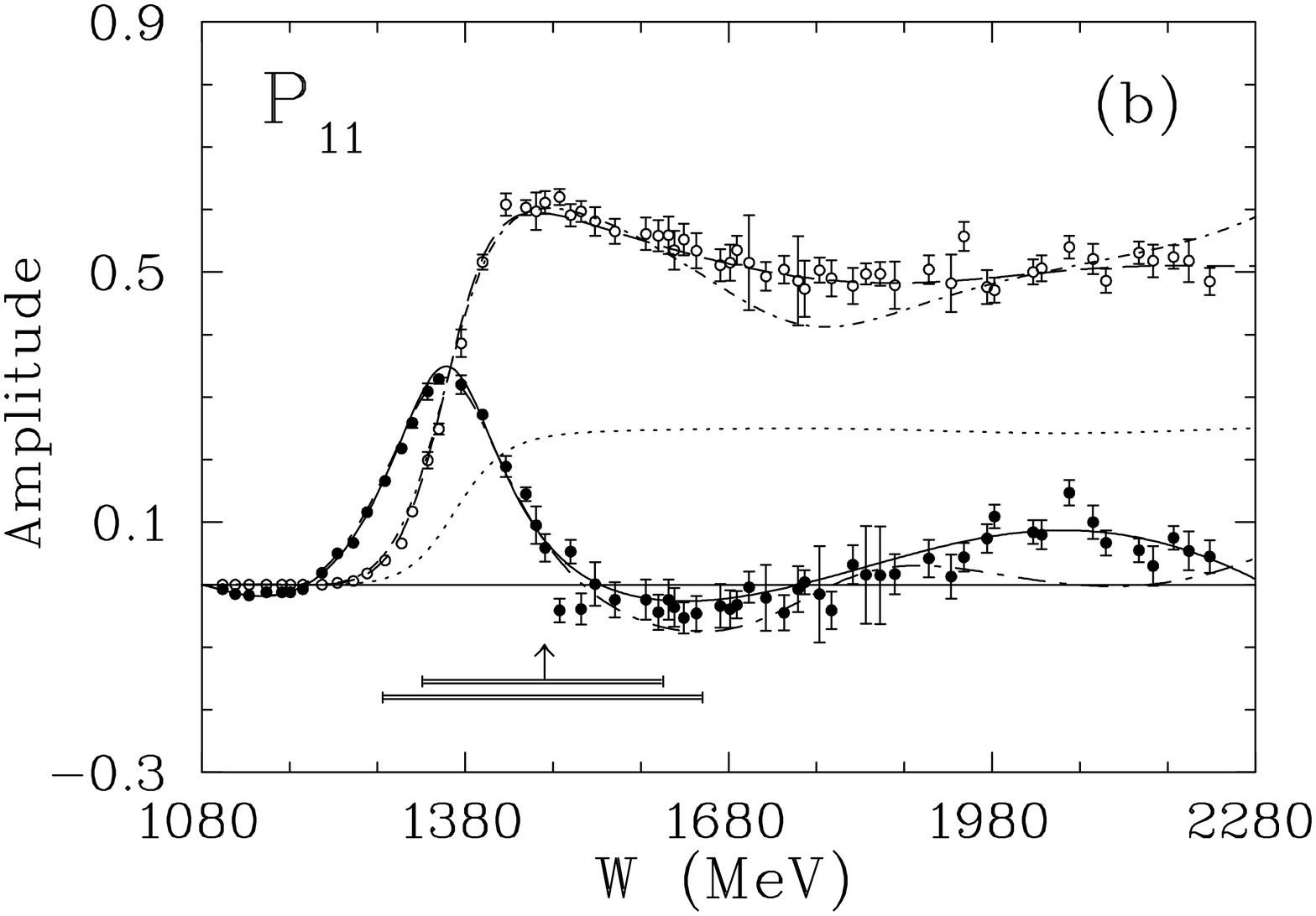}\hfill
\includegraphics[height=0.4\textwidth, angle=90]{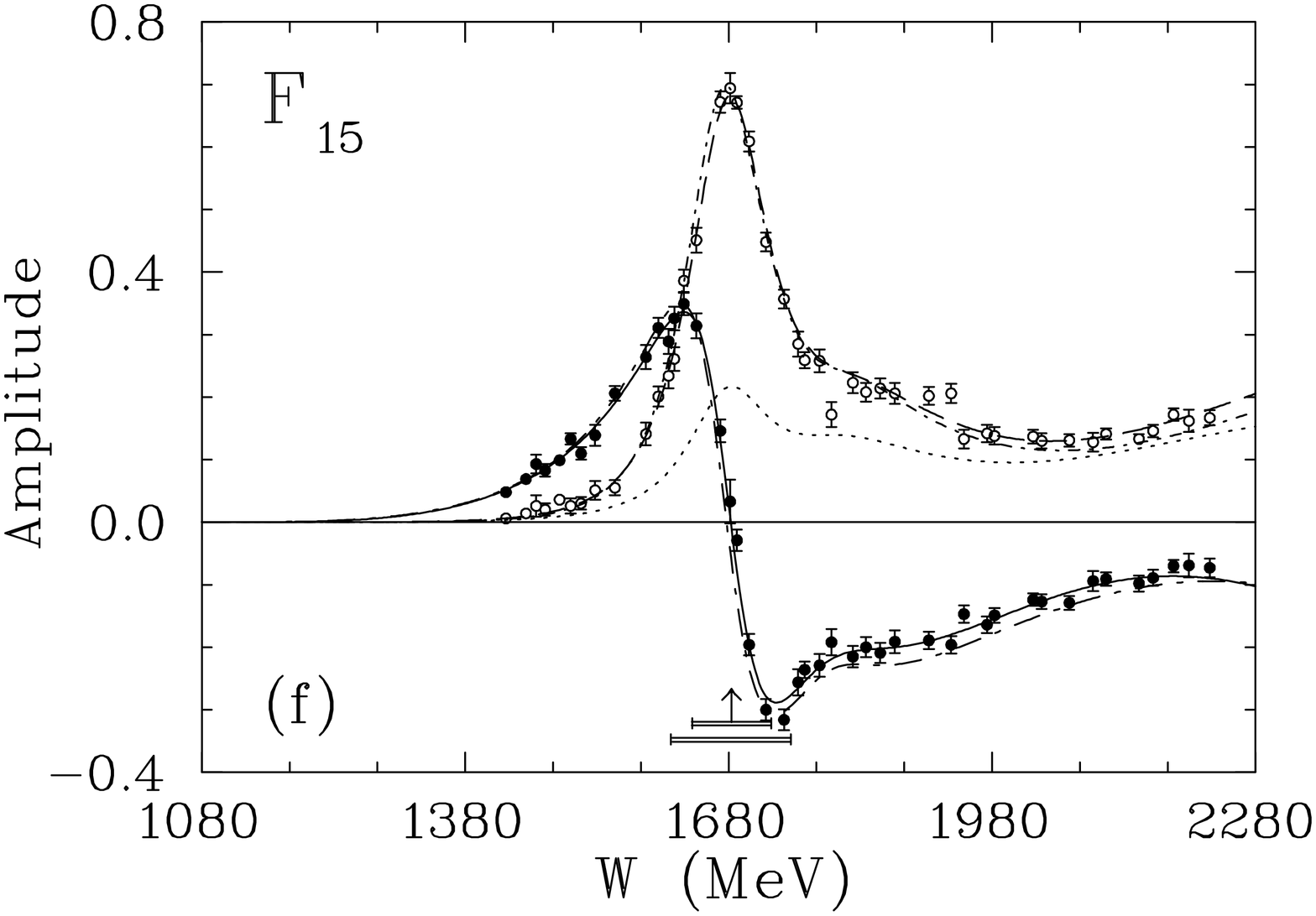}
\includegraphics[height=0.4\textwidth, angle=90]{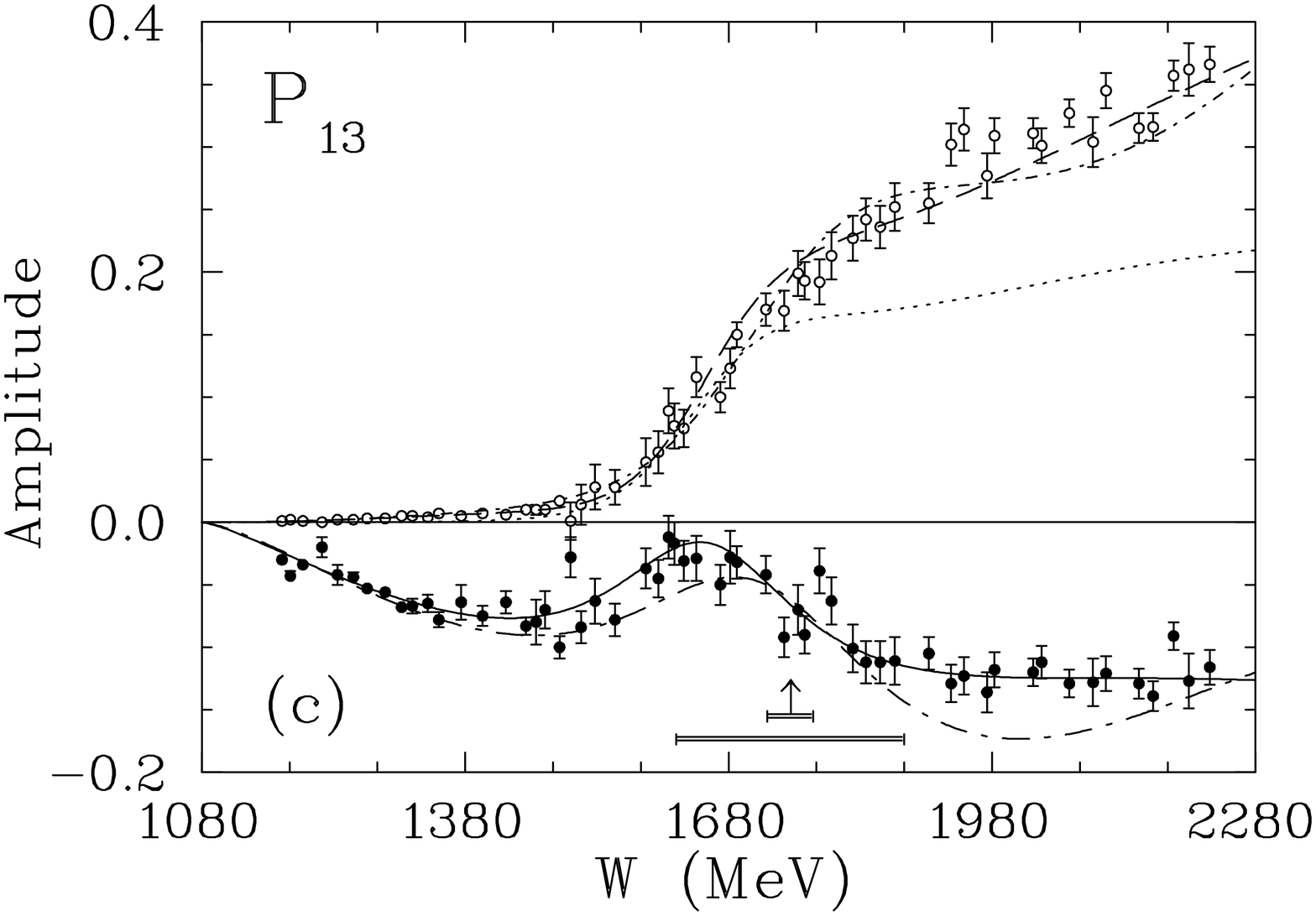}\hfill
\includegraphics[height=0.4\textwidth, angle=90]{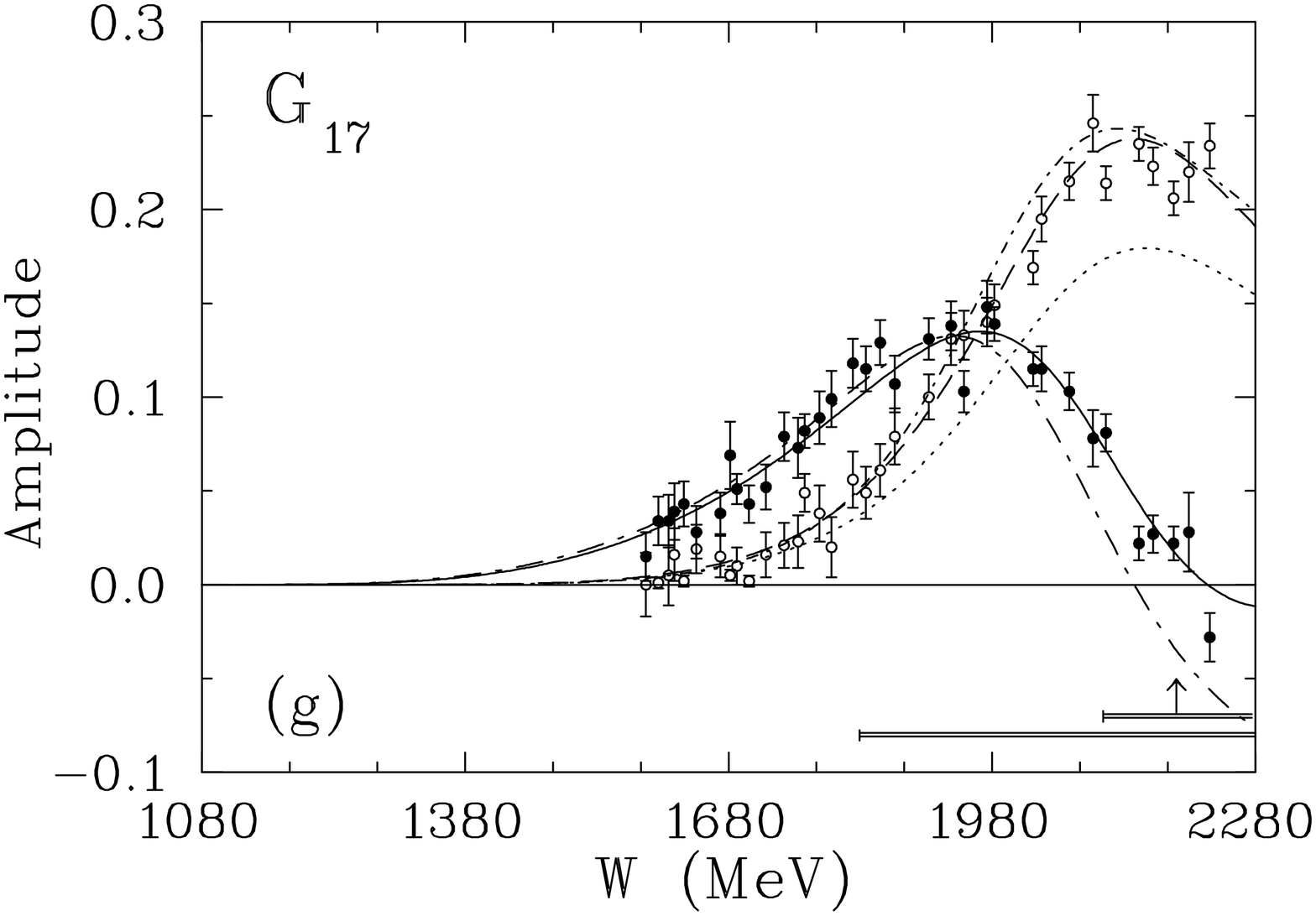}
\includegraphics[height=0.4\textwidth, angle=90]{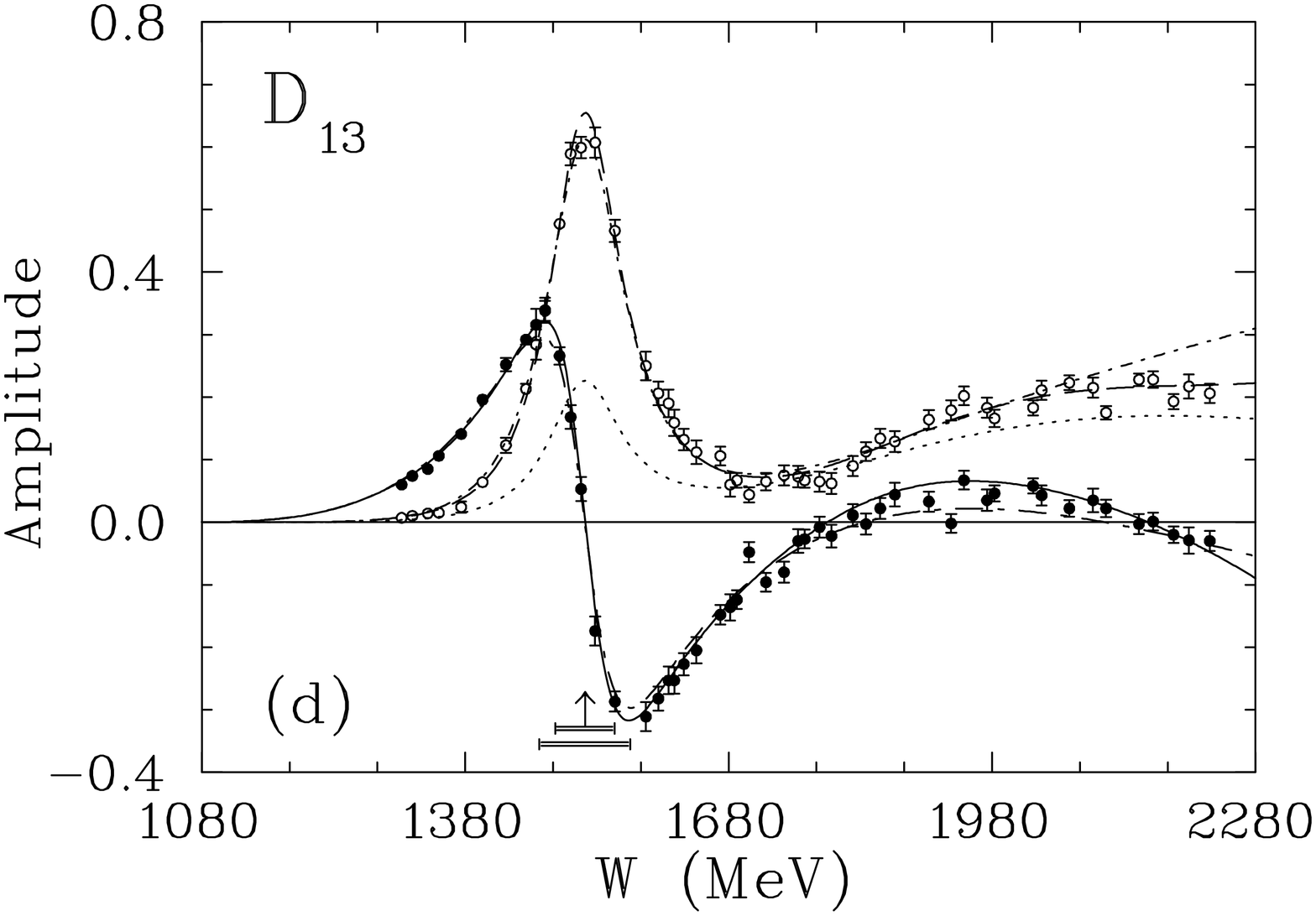}\hfill
\includegraphics[height=0.4\textwidth, angle=90]{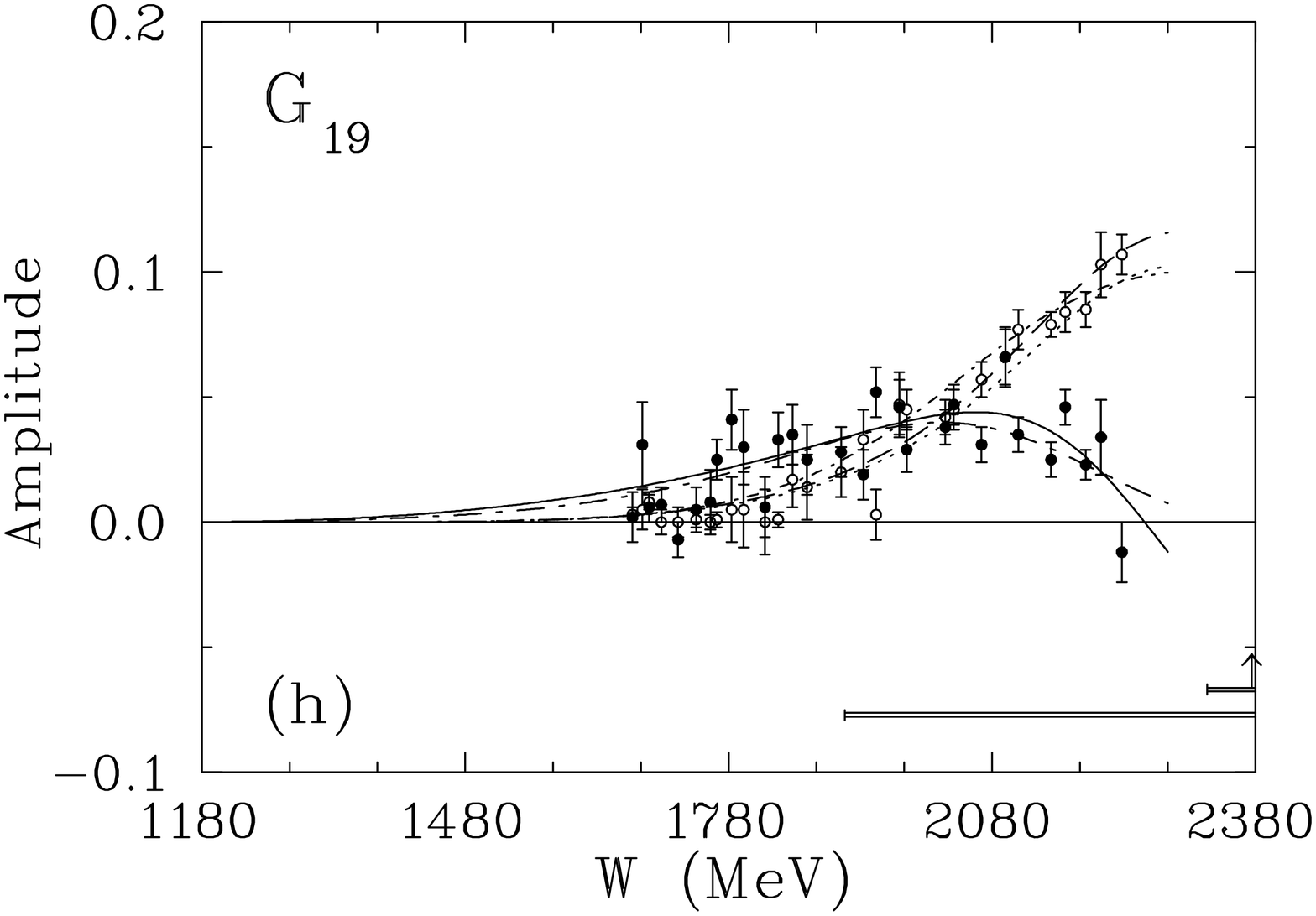} 
}\caption{Isospin $1/2$ partial-wave amplitudes (L$_{2I, 2J}$)
      from $T_{\pi}$ = 0 to 2.1~GeV.  Solid (dashed) curves
      give the real (imaginary) parts of amplitudes  
      corresponding to the FA02 solution.  The real
      (imaginary) parts of single-energy solutions are
      plotted as filled (open) circles.  The previous
      SM95 solution~\protect\cite{sm95} is plotted with
      long dash-dotted (real part) and short dash-dotted
      (imaginary part) lines.  The dotted curve gives the
      unitarity limit ($ImT - T^{\ast}T$) from FA02.  All
      amplitudes are dimensionless.  Vertical arrows
      indicate $W_R$ and horizontal bars show full
      $\Gamma$/2 and partial widths for $\Gamma_{\pi N}$
      associated with the FA02 results presented in Table~
      \protect\ref{tbl4}.
      (a) $S_{11}$, (b) $P_{11}$, (c) $P_{13}$, (d) $D_{13}$,
      (e) $D_{15}$, (f) $F_{15}$, (g) $G_{17}$, and
      (h) $G_{19}$. \label{fig:g6}}
\end{figure}

\begin{figure}[th]
\centering{
\includegraphics[height=0.4\textwidth, angle=90]{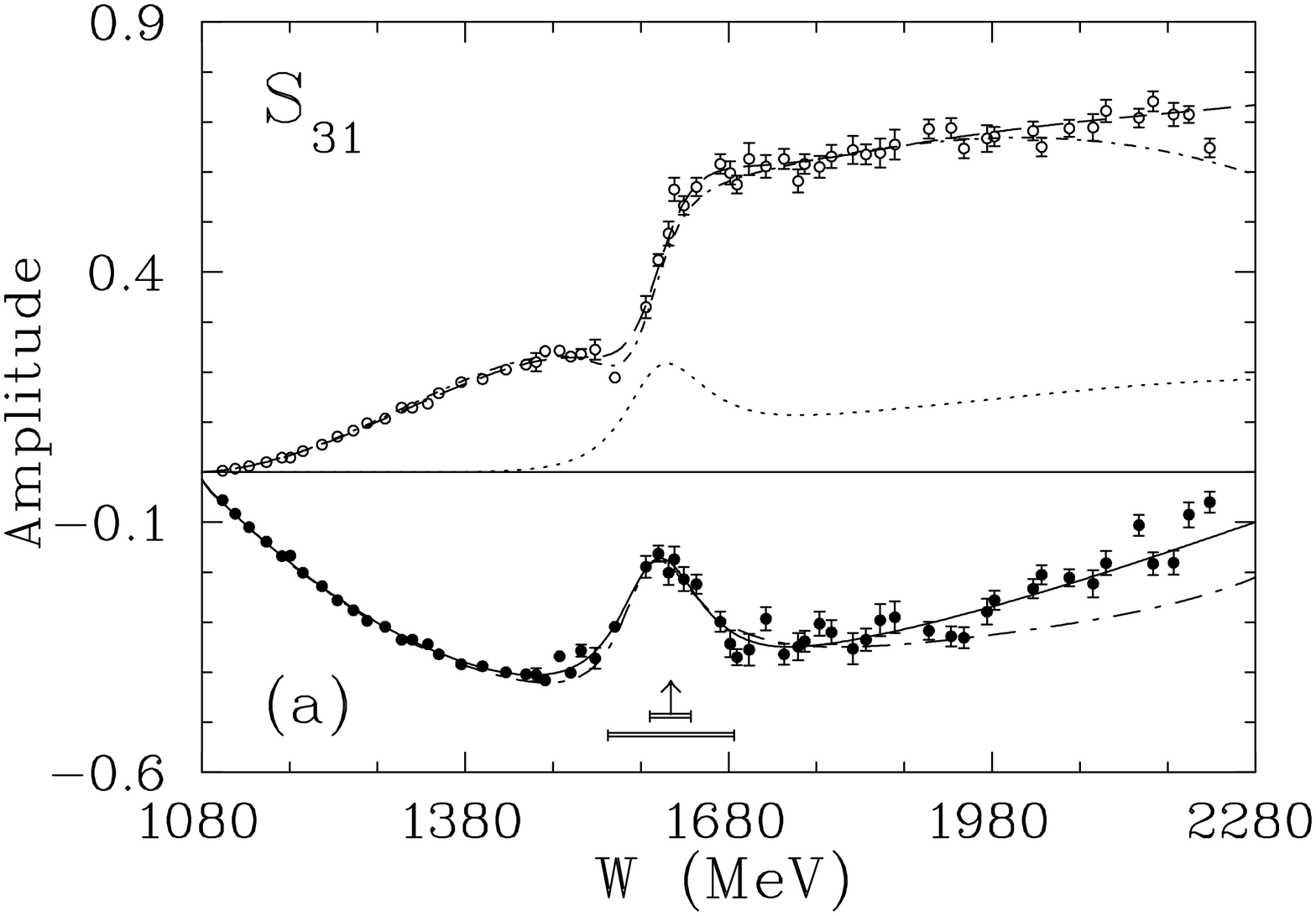}\hfill
\includegraphics[height=0.4\textwidth, angle=90]{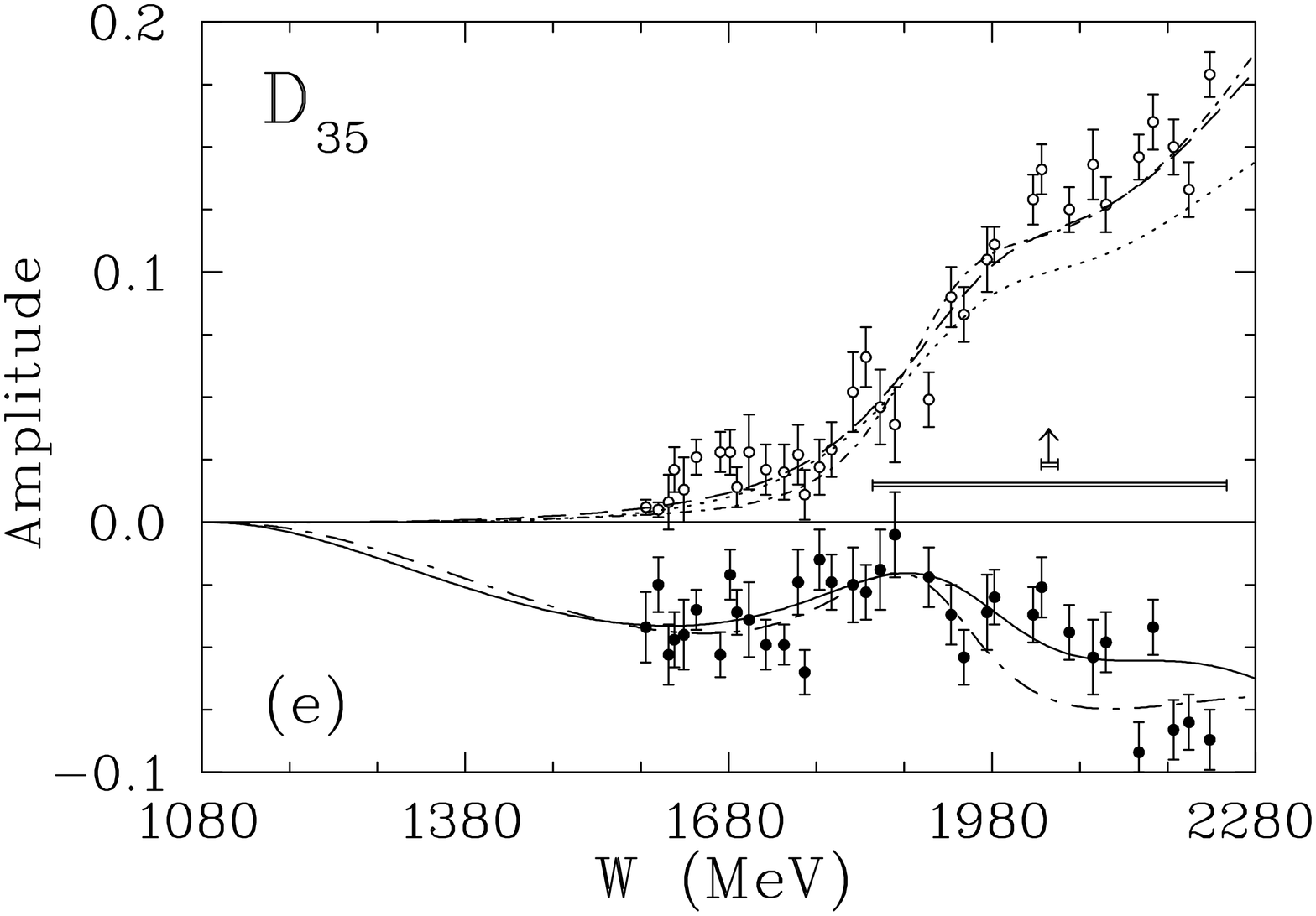}
\includegraphics[height=0.4\textwidth, angle=90]{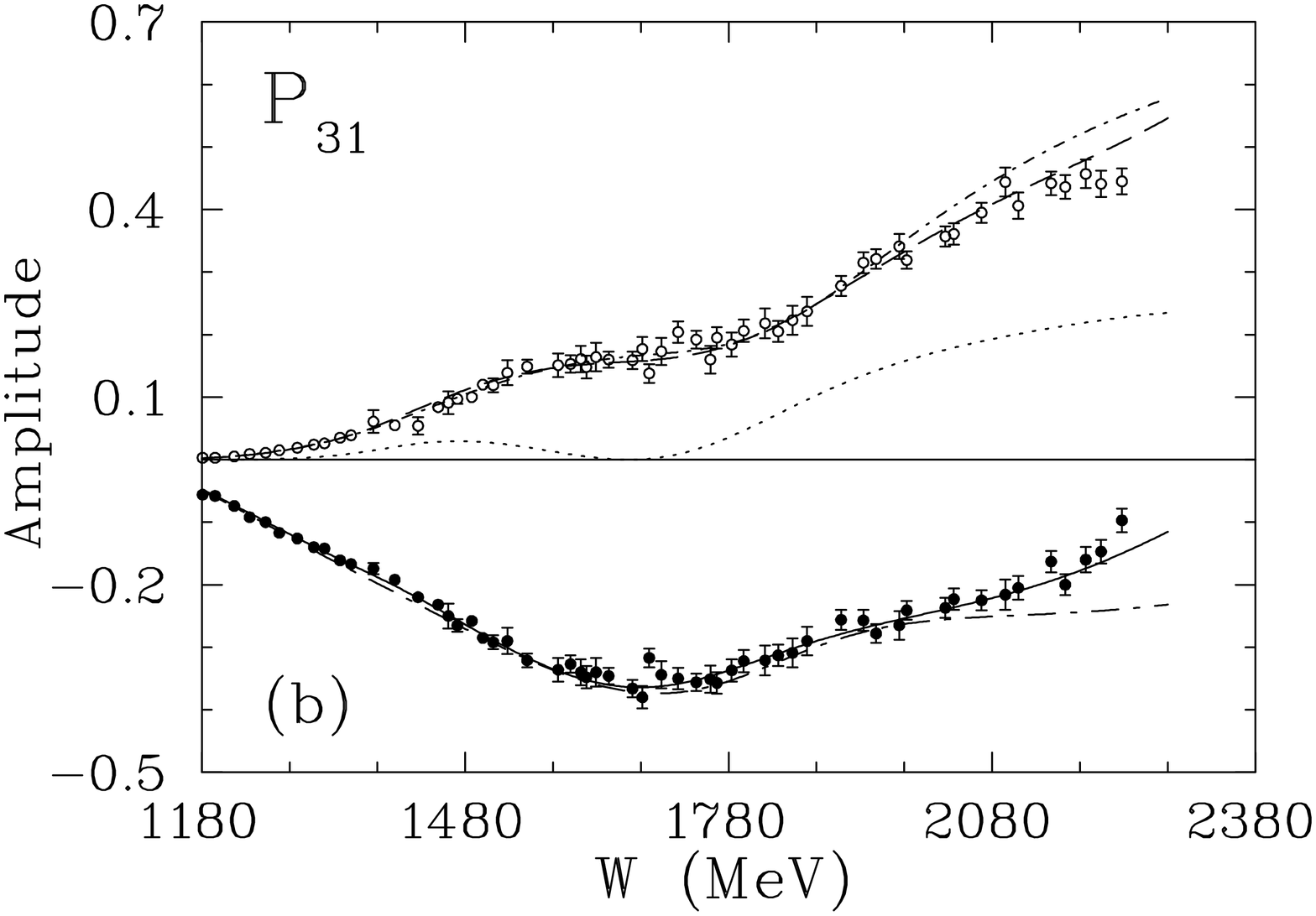}\hfill
\includegraphics[height=0.4\textwidth, angle=90]{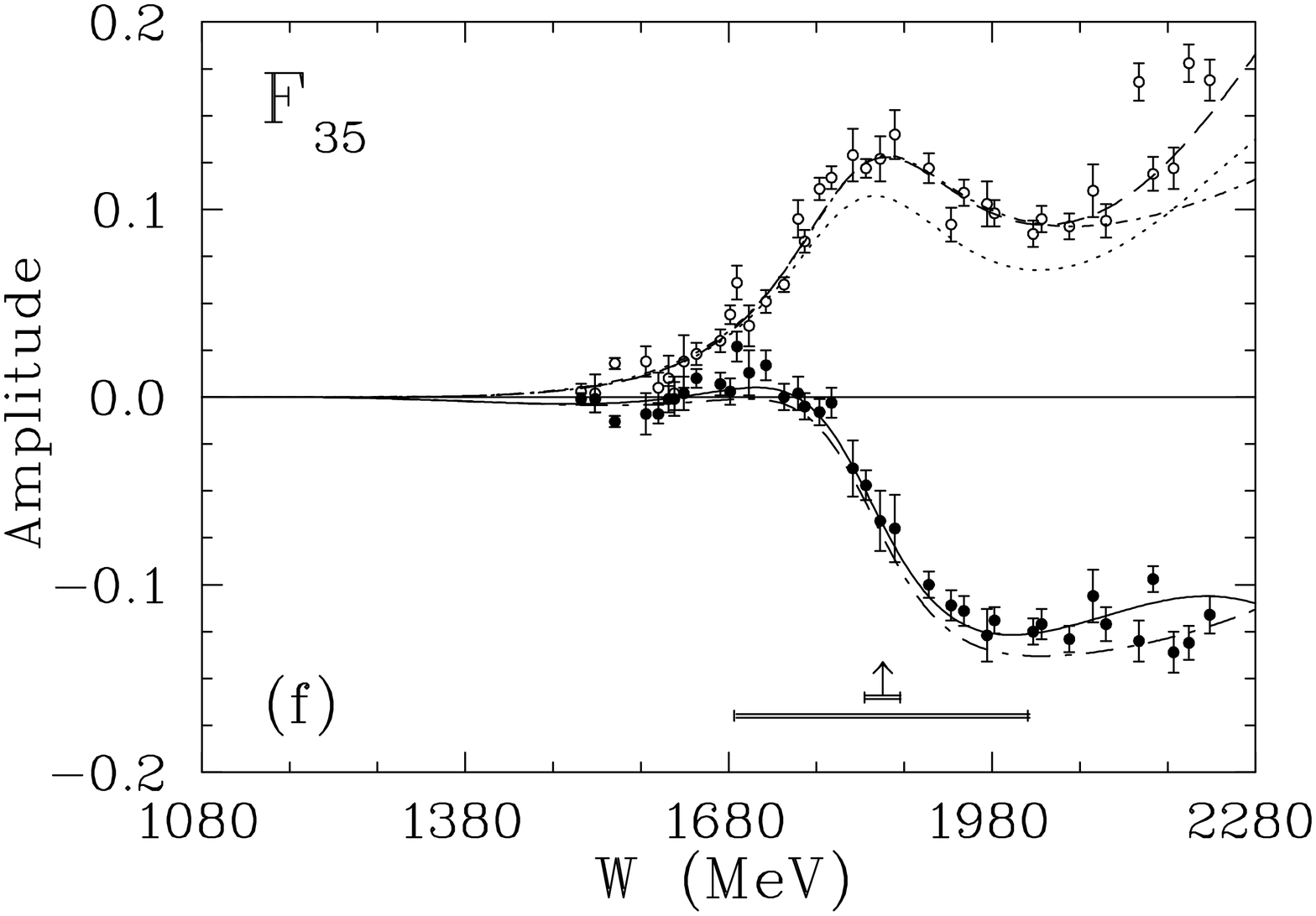}
\includegraphics[height=0.4\textwidth, angle=90]{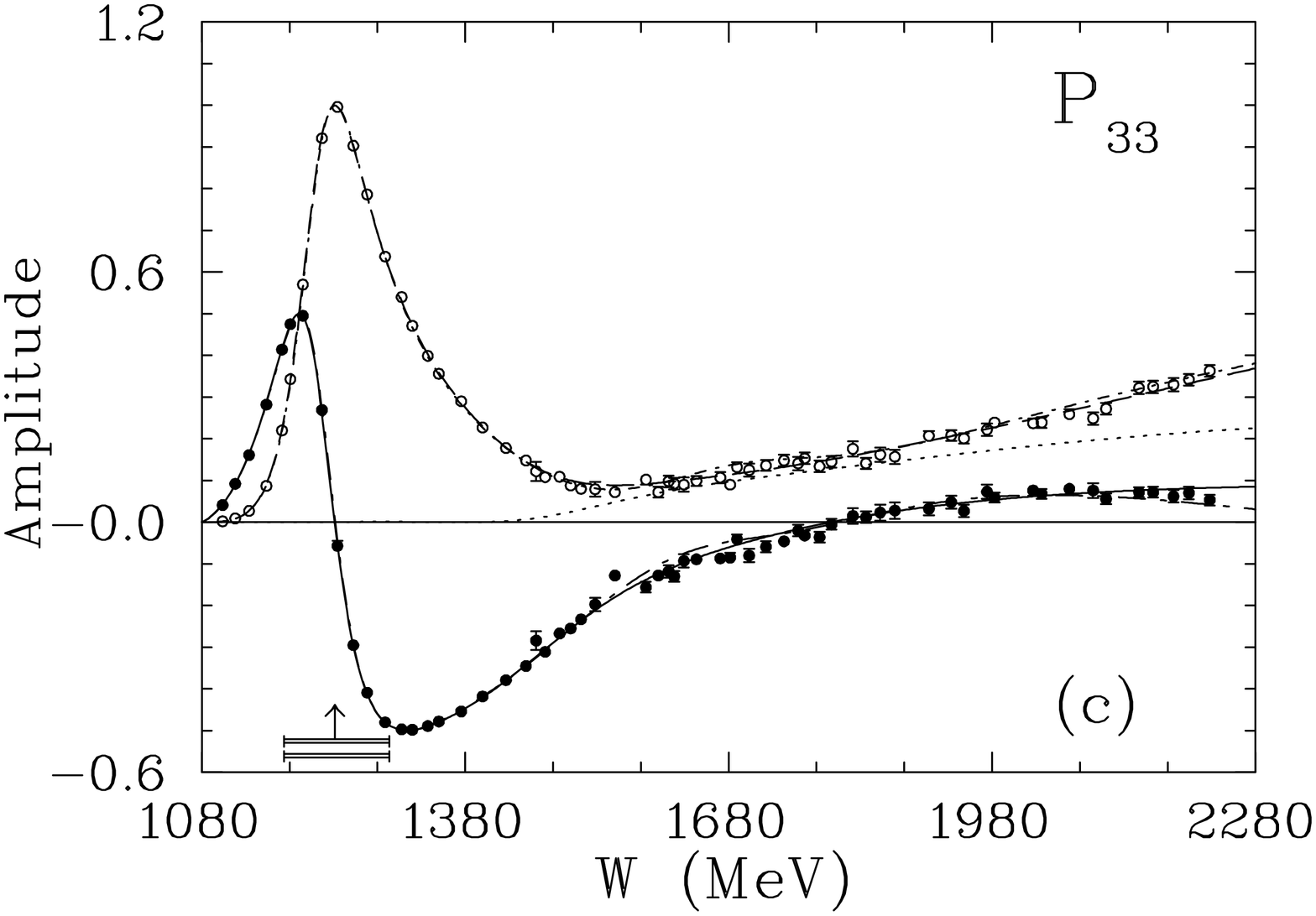}\hfill
\includegraphics[height=0.4\textwidth, angle=90]{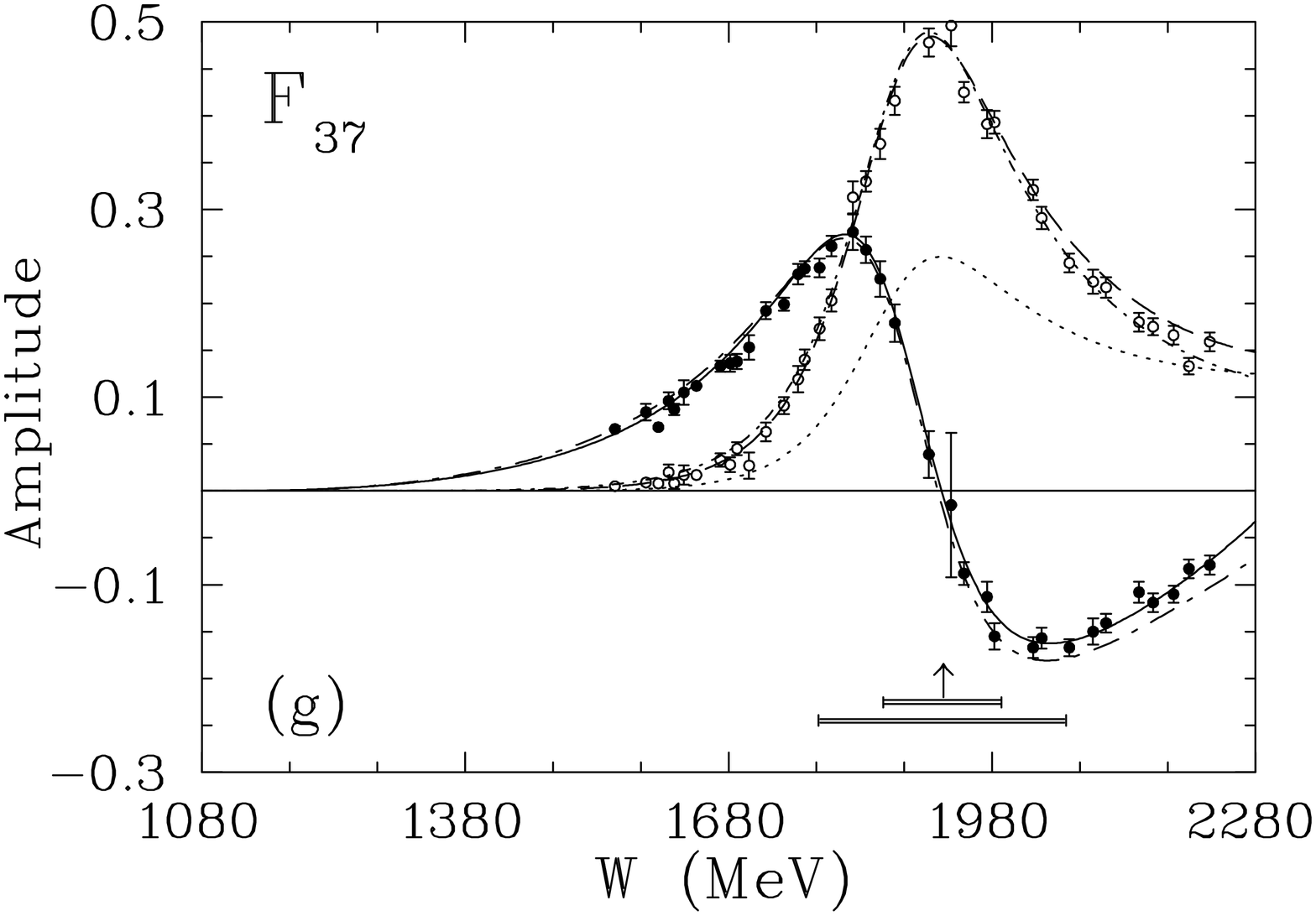}
\includegraphics[height=0.4\textwidth, angle=90]{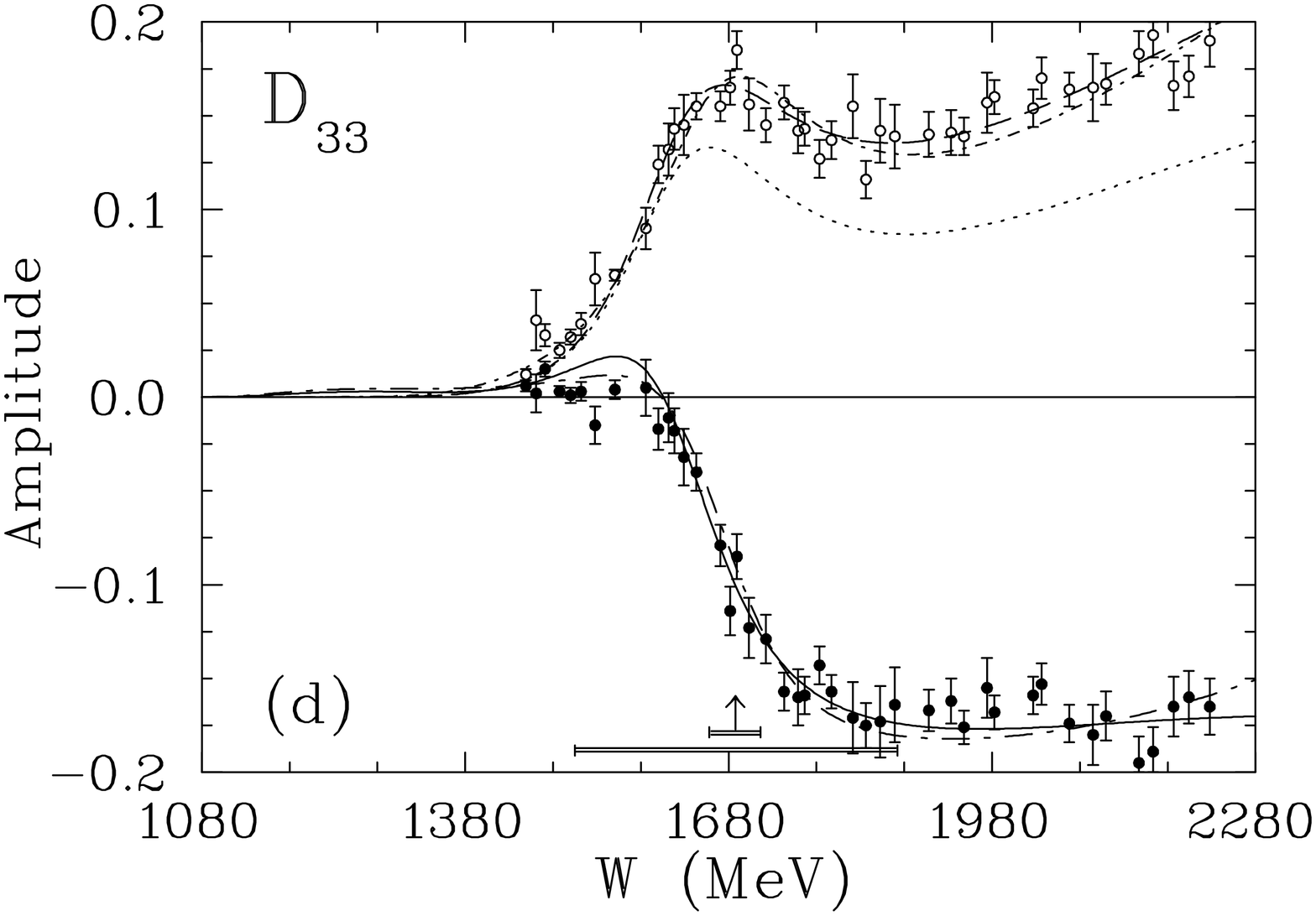}\hfill
}\caption{Isospin $3/2$ partial-wave amplitudes (L$_{2I, 2J}$)
      from $T_{\pi}$ = 0 to 2.1~GeV.  Notation as in Fig.~
      \protect\ref{fig:g6}.  Vertical arrows indicate $W_R$
      and horizontal bars show full $\Gamma$/2 and partial
      widths for $\Gamma_{\pi N}$ associated with the FA02
      results presented in Table~\protect\ref{tbl5}.
      (a) $S_{31}$, (b) $P_{31}$, (c) $P_{33}$, (d) $D_{33}$,
      (e) $D_{35}$, (f) $F_{35}$, and (g) $F_{37}$.
      \label{fig:g7}}
\end{figure}

Figs.~\ref{fig:g6} and~\ref{fig:g7} compare the 
energy-dependent fits FA02 and SM95 over the full 
energy region.  Partial waves with $l < 6$ are 
displayed, whereas the analysis has fitted waves up to 
$l = 6$.  Deviations from SM95 are visible in amplitudes $S$ and 
$P$ for $I = 1/2$, and the $D$ waves, mainly near the end point of 
our analysis.  Considering the complicated structure associated 
with an opening $\eta$n channel and N(1535) resonance, the S$_{11}$ 
partial-wave shows remarkably little deviation from the earlier 
SM95 fit. 

\subsection{Resonance Parameter Extraction}
\label{sec:res}

\begin{figure}[th]
\centering{
\includegraphics[height=0.4\textwidth, angle=90]{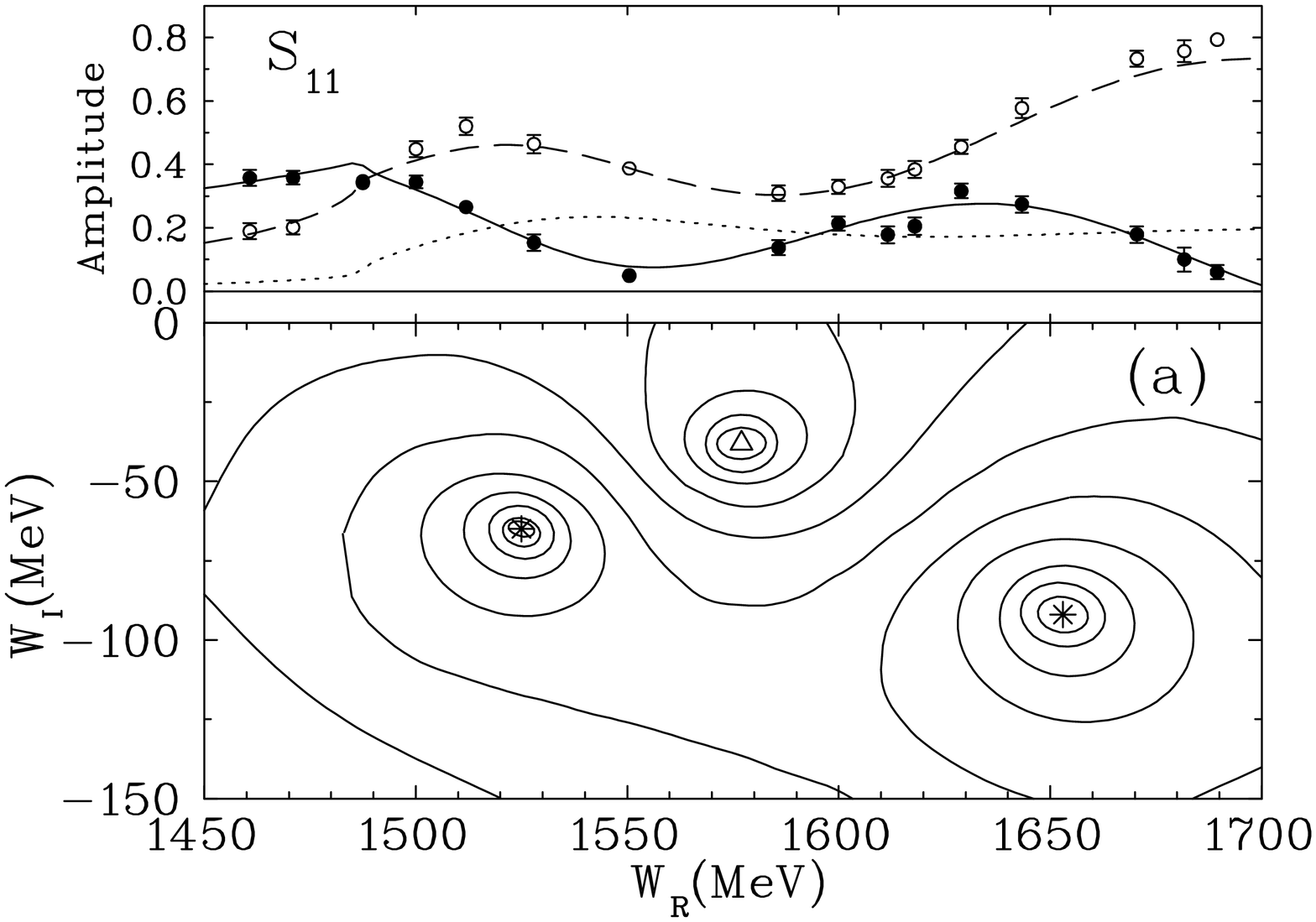}\hfill
\includegraphics[height=0.4\textwidth, angle=90]{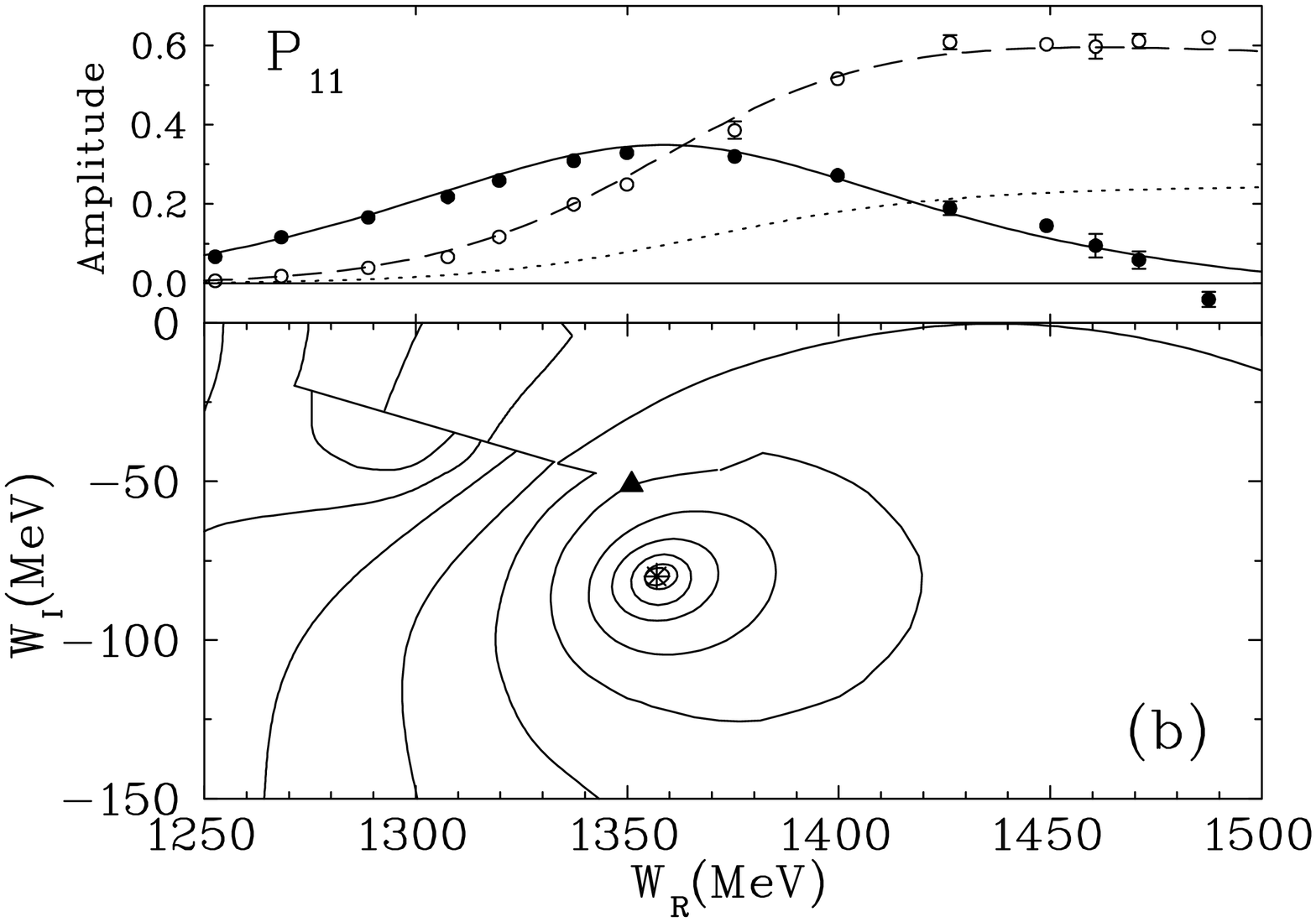}\hfill
\includegraphics[height=0.4\textwidth, angle=90]{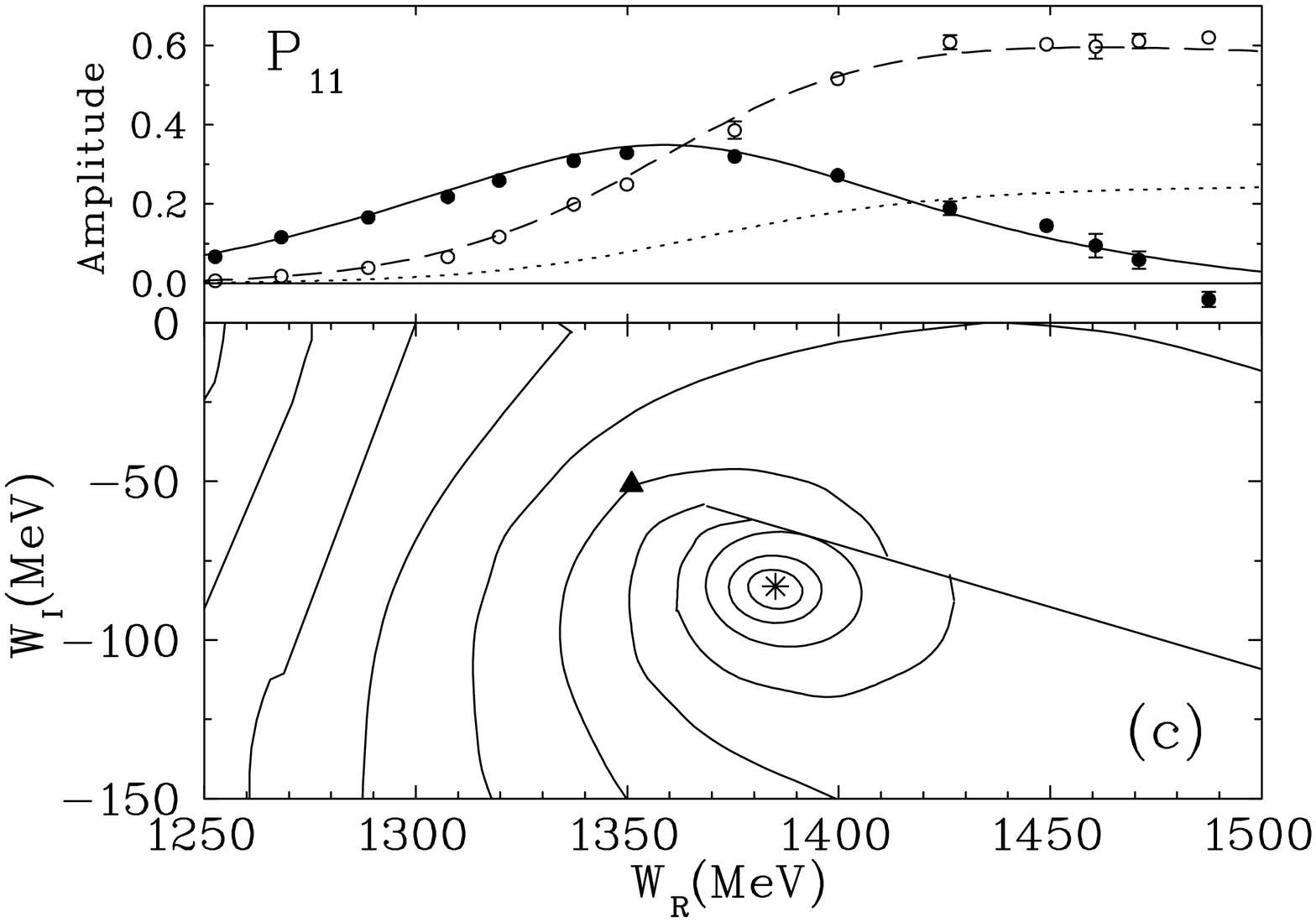}
}\caption{Contour plot of $ln|T|^2$ displaying complex plane
          structures.
          (a) S$_{11}$.  Poles: W$_P$ = 1526 - i65~MeV   
          and W$_P$ = 1653 - i91~MeV, shown as stars.
          Zero: W$_Z$ = 1578 - i38~MeV, represented as
          an open triangle;
          (b) P$_{11}$, first sheet.  Pole: W$_P$ =
          1357 - i80~MeV, shown as a star; and
          (c) P$_{11}$, second sheet.  Pole: W$_P$ =
          1385 - i83~MeV, shown as a star.  The
          branch point for P$_{11}$, 1349 - i50~MeV,
          is represented as a solid triangle.
          The $\pi\Delta$ branching cut in (b) and (c)
          is shown as a solid line. \label{fig:g8}}
\end{figure}

The resonance spectrum from our fit has been extracted  
using three different methods. Poles and zeros have 
been found by continuing into the complex energy plane.  
These are compiled in Tables~\ref{tbl6},~\ref{tbl7}, and~
\ref{tbl8} along with the modulus and phase of all pole 
residues.  The complicated interplay of poles and zeros 
is displayed in Fig.~\ref{fig:g8} for the S$_{11}$ and 
P$_{11}$ partial waves. Clearly, for these partial 
waves and their associated resonances, a simple 
Breit-Wigner parameterization should be avoided. An error
in the routine used to determine pole residues for SM95
was found~\cite{res_err} and corrected. The most significant
change is seen in the $\Delta (1232)$ residue, which now
agrees with most previous determinations.

\begin{figure}[th]
\centering{
\includegraphics[height=0.4\textwidth, angle=90]{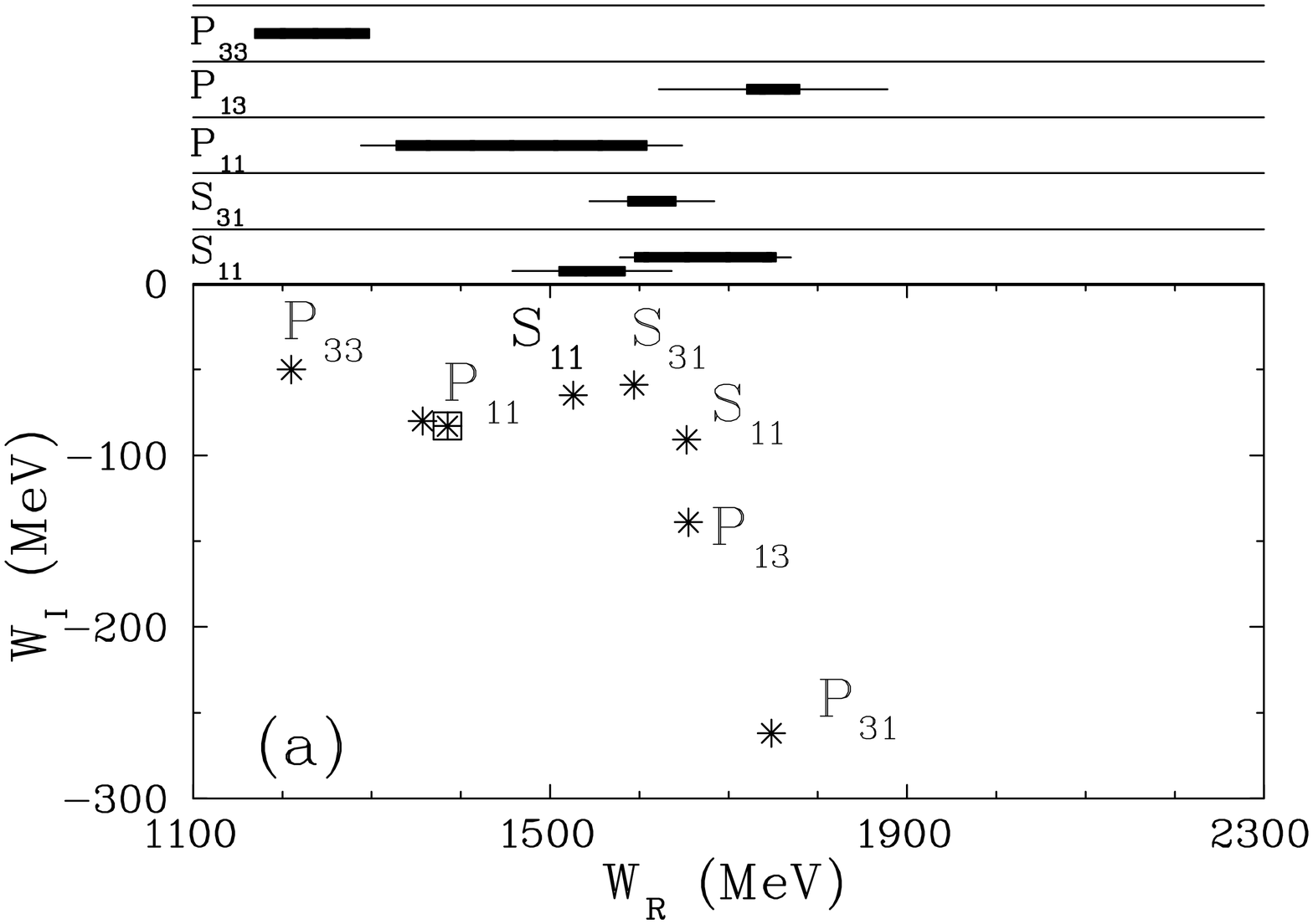}\hfill
\includegraphics[height=0.4\textwidth, angle=90]{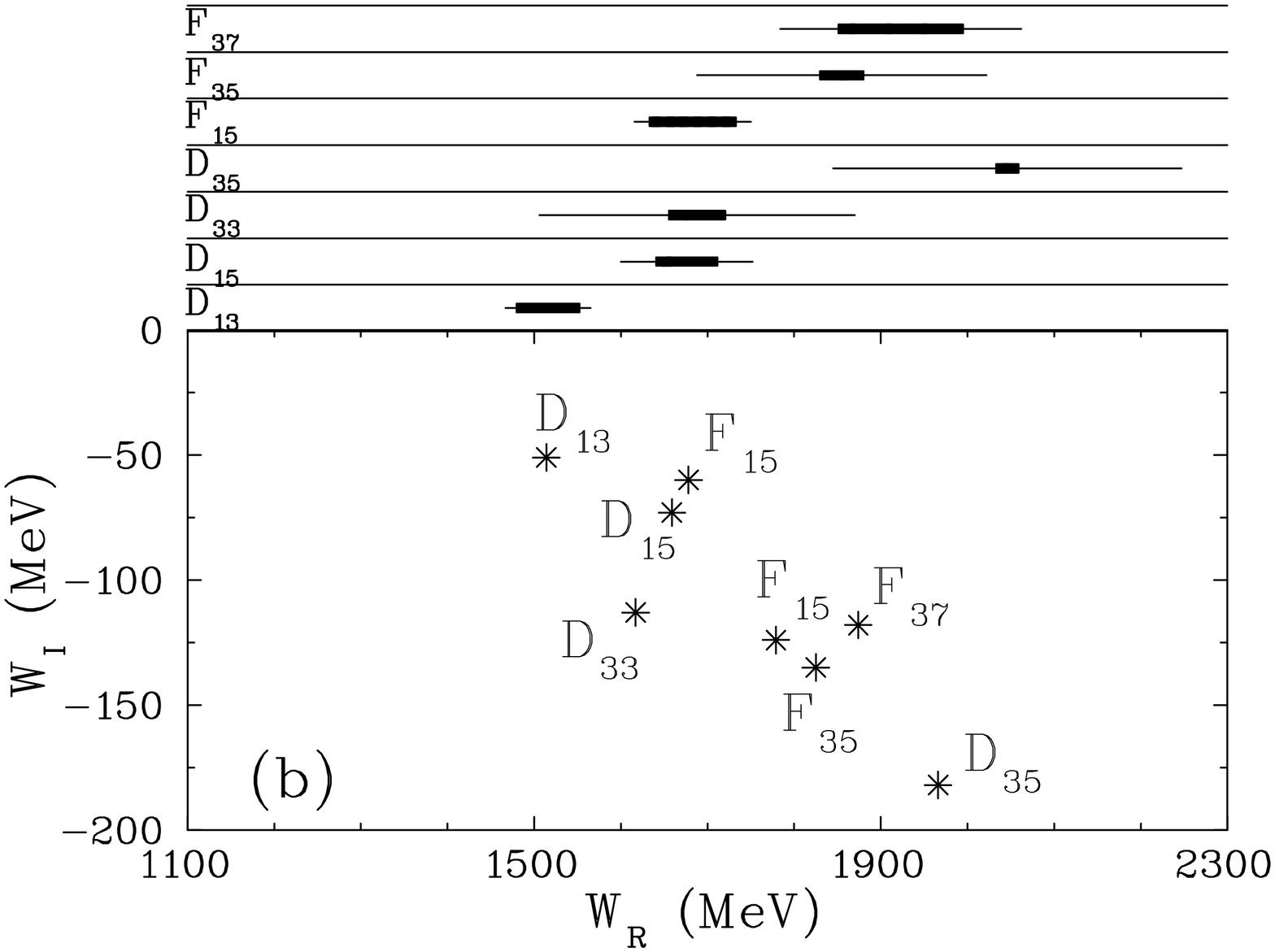}
}\caption{Comparison of complex plane (bottom panel)
          and Breit-Wigner (top panel) fits for
          resonances found in the FA02 solution.
          Plotted are the result for (a) S- and
          P-wave resonances and (b) D- and F-wave
          resonances.  Complex plane poles are shown
          as stars (the boxed star denotes a
          second-sheet pole).  $W_R$ and $W_I$ give
          real and imaginary parts of the
          center-of-mass energy.  The full ($\pi N$
          partial) widths are denoted by narrow (wide)
          bars for each resonance. \label{fig:g9}}
\end{figure}

More commonly used, though more model-dependent, are 
Breit-Wigner parameters for the resonances. Here, as 
for the SM95 fit, a Breit-Wigner plus background 
contribution was initially fitted to the single-energy 
amplitudes, over varying ranges of energy.  However, 
given the scatter often found in the single-energy 
values, we have rejected this method in favor of 
another.  In Tables~\ref{tbl4} and~\ref{tbl5}, 
we give resonance parameters obtained from a 
Breit-Wigner plus background (using a product S-matrix 
approach, $S = S_R S_B$) representation applied 
directly to the data.  Here, $S_R$ = 1 + 2 i $T_R$, with 
\begin{eqnarray}
T_R =  {{ \Gamma_{\pi} / 2} \over { W_R - W -i \left( \Gamma_{\pi} 
/2 + \Gamma_I / 2 \right) }}.
\label{5}\end{eqnarray}
The total width is given by $\Gamma = \Gamma_{\pi} + \Gamma_I$,
where
\begin{eqnarray}
\Gamma_{\pi} & = & \rho_{\pi} \Gamma R ,
\label{6}\end{eqnarray}
\begin{eqnarray}
\Gamma_I     & = & \rho_i \Gamma ( 1 - R) , 
\nonumber\end{eqnarray}
with $R$ being the branching fraction to $\pi$N.  The background 
T-matrix is given in terms of a K-matrix, $T_B = K_B ( 1 - i 
K_B)^{-1}$, with $K_B =  A + B (W - W_R)  + i C$. Only one
background parameter was necessary for the $I = 3/2$ waves.  Data were 
then fitted using this representation for a particular resonant 
partial wave. The remaining waves were fixed to values found in 
the full global analysis.  Values for the background parameters, 
energy ranges over which fits were performed, and $\chi^2$ 
comparisons are given in Tables~\ref{tbl9} and~\ref{tbl10}.  
In Fig.~\ref{fig:g9}, S-, P-, D-, and F-wave resonances are 
compared in terms of their pole positions versus a fitted 
Breit-Wigner mass and width. 

In Table~\ref{tbl4}, results for the N(1535) and N(1650) 
resonances are significantly different than those reported in the
SM95 analysis. The widths of both states are now closer to their
RPP averages. The N(1650) continues to show a nearly elastic
behavior in the FA02 solution. This structure also appears to 
be difficult to parameterize, requiring the most background 
parameters in the present representation. Breit-Wigner parameters
are absent for the $\Delta (1920)$ as this fit produces a mass
about 500 MeV greater than suggested by the pole position. 

In Table~\ref{tbl11}, we give resonance parameters determined from
a fit to data without DR constraints. In most cases, there is 
reasonable agreement with the constrained fit. An exceptions 
appears to be the $P_{13}(1720)$, which has a relatively weak
coupling to $\pi N$ states, and has an atypical resonance 
signature.

\subsection{Charge-Symmetry Violation in $\pi N$ 
Scattering}
\label{sec:csv}

The issue of charge symmetry violation (CSV) is fundamental to 
our understanding of hadronic interactions, and many experimental 
and theoretical studies have addressed this issue (see review~
\cite{mill}).  In the framework of QCD, CSV arises from the mass 
difference between the $u$ and $d$ quarks.  The other principal 
cause for CSV comes from the electromagnetic interaction, as
described in Sec.~\ref{sec:clmb}.  Weinberg~
\cite{wein} pointed out that the effective chiral $\pi$N Lagrangian, 
coming from QCD, contains a term which violates charge symmetry (see 
also, a recent review by Meissner~\cite{meis}).  Thus, not only are 
there kinematic reasons for CSV due to the mass differences within 
baryon multiplets, but direct CSV effects should exist as well.  
Recent analyzes~\cite{csv,matt} of the triangle identity in
low-energy $\pi$N scattering (below 70~MeV)
data have found some indications for significant (6-7\%) 
direct CSV effects in the 
strong-interaction sector~\cite{csv,matt}. 
Fettes and Meissner~\cite{fett00} have found a somewhat 
smaller violation ($\sim$2.5\%) in a third order chiral
perturbation theory calculation without electromagnetic effects, and
even smaller violations once the latter were included~\cite{fett01}.

Because our formalism does not employ CSV beyond the Coulomb 
interaction, which we take into account in our treatment of the 
database, we have generated a test fit (X370) of the full database 
excluding charge-exchange data. Differences between FA02 and X370
are therefore signal incompatibilities between the elastic and
charge-exchange data, which could be an indication of CSV.

In Table~\ref{tbl12}, we compare 
the FA02 and X370 solutions.  For the energy range associated with 
meson factories (below 500~MeV), there is little difference between 
FA02 and X370.  Above 500~MeV, a significant $\chi^2$ contribution 
comes from the old Rutherford High Energy Laboratory (RHEL) 
$\pi^+p$ P measurements where there are 1976 P data from 480 to 
2080~MeV~\cite{ma75} favoring by $\chi^2\sim$200 the fit X370 versus 
FA02.  Conversely, for charge-exchange $d\sigma 
/d\Omega$ two old RHEL sets (416 cross sections from 900 to 2050~MeV~
\cite{ne72}, 156 backward cross sections from 480 to 870~MeV~
\cite{de75}, and one KEK set (72 cross sections from 1830 to 
2040~MeV~\cite{su87}) are more poorly represented, by $\chi^2\sim$1900, 
in X370 versus FA02.  
At very low energies, there exist rather few charge-exchange data,
and almost no polarization data, so a precise test of the 
Fettes and Meissner~\cite{fett00,fett01} prediction is not possible.
As a result, we cannot claim 
any compelling evidence for sizable CSV effects in $\pi$N 
elastic scattering.  This is consistent with recent findings
for $\pi 2N$ and $\pi 3N$ systems~\cite{itepgw}.

\subsection{Dispersion Relation Parameters}
\label{sec:drResults}

\begin{figure}[th]
\centering{
\includegraphics[height=0.4\textwidth, angle=90]{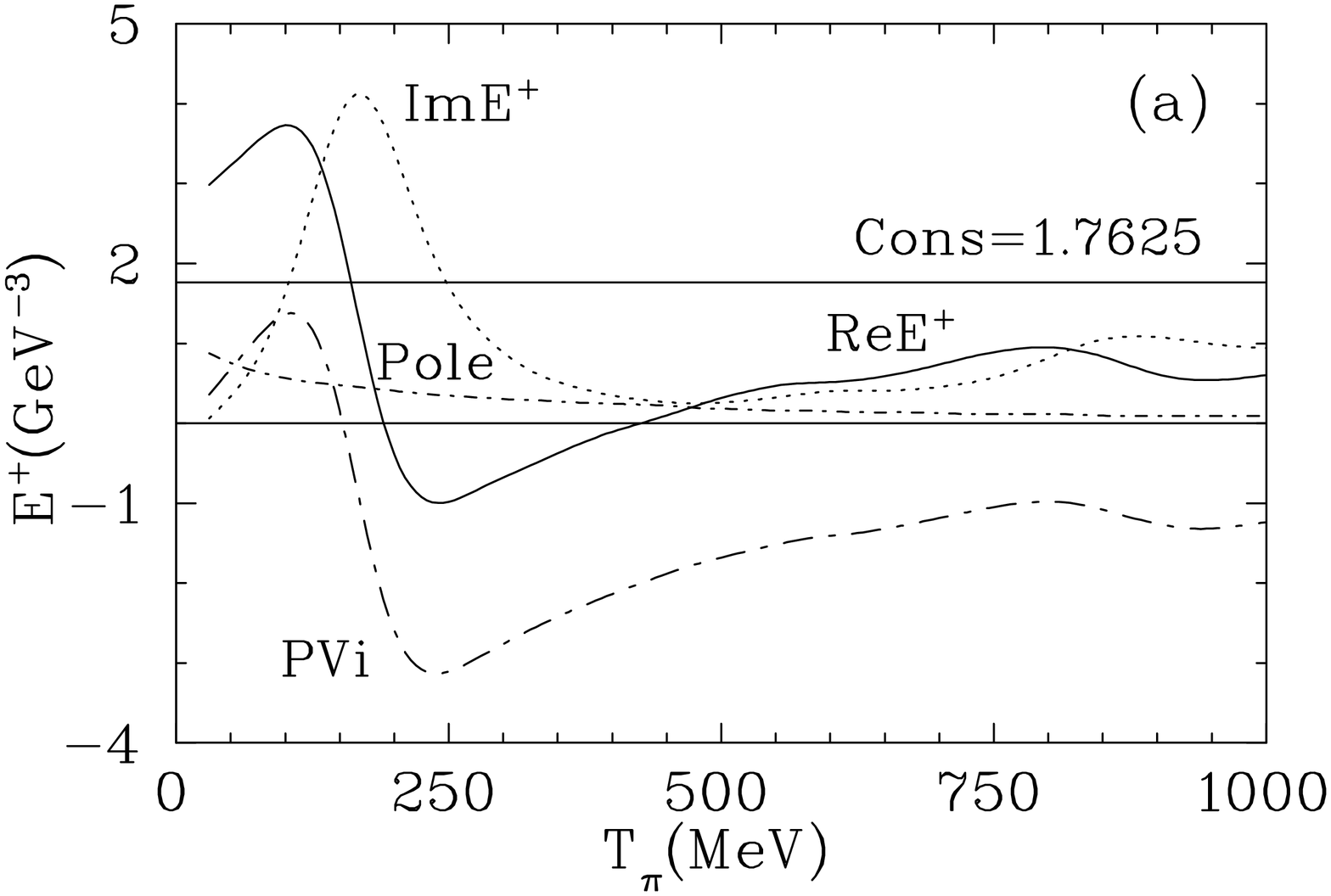}\hfill
\includegraphics[height=0.4\textwidth, angle=90]{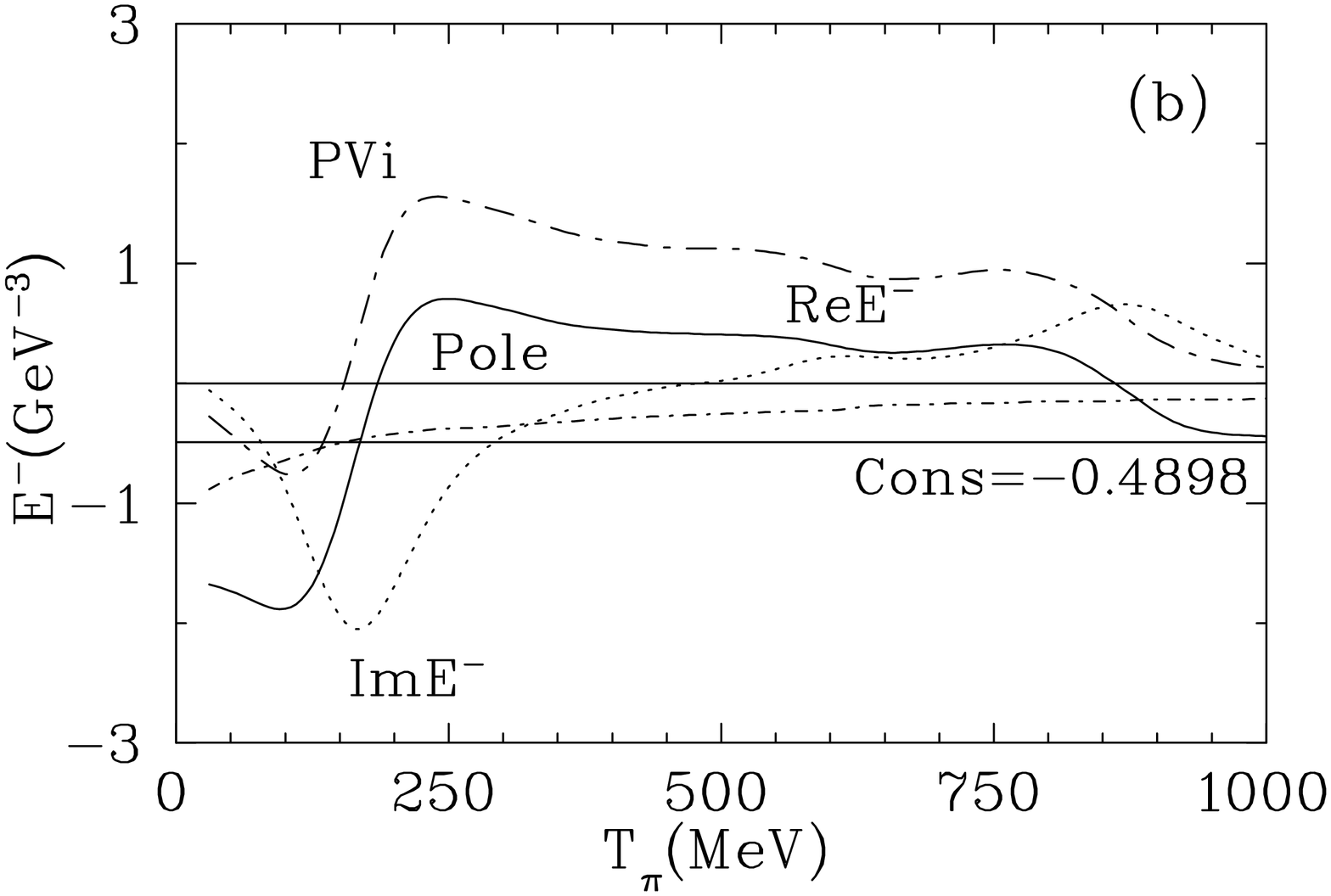}
}\caption{Forward derivative (a) E$^+$($\omega$) and (b)
          E$^-$($\omega$) dispersion relations.  ReE 
          (ImE) are plotted by
          solid (dashed) lines, the principal value
          integral, PVi, by dash-dotted lines, and the
          pole term by short-dash-dotted lines.  The
          respective subtraction constants are shown 
          as horizontal solid lines. \label{fig:g4}}
\end{figure}

\begin{figure}[th]
\centering{
\includegraphics[height=0.7\textwidth, angle=90]{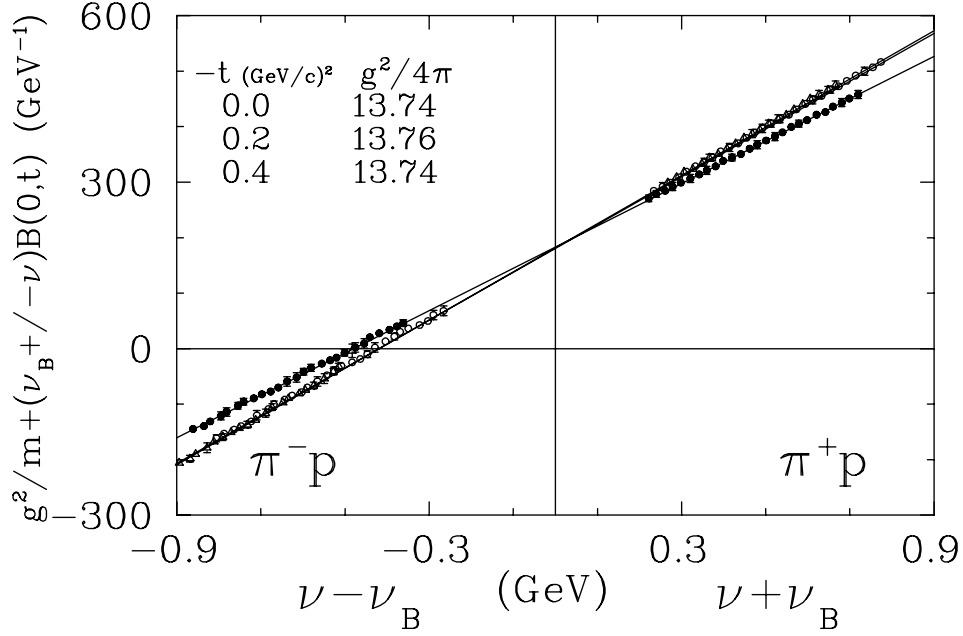}
}\caption{The H\"uper dispersion relation~
          \protect\cite{hoehler} plotted for several
          values of four-momentum transfer from 0 to 
          $-0.4~(GeV/c)^2$.  The y-intercept yields
          the coupling constant $g^2/4\pi$.  Lines
          represent the least-square averages of 
          individual values. \label{fig:g10}}
\end{figure}

\begin{figure}[th]
\centering{
\includegraphics[height=0.7\textwidth, angle=90]{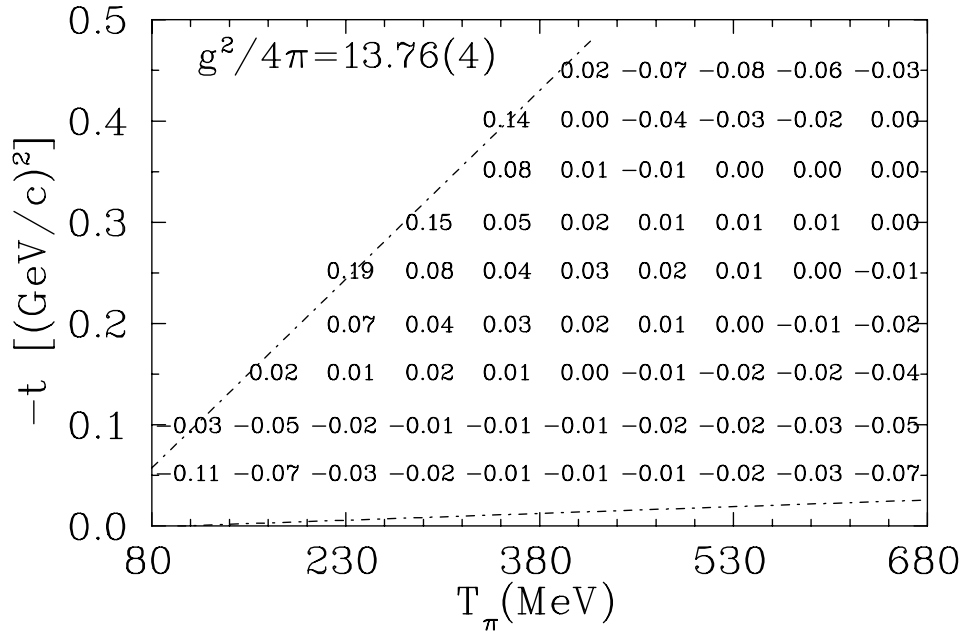}
}\caption{Deviations from the mean value of $g^2/4\pi$ for
          FA02 from the $B^+ (\nu , t)$ DR evaluated
          over a grid of T$_{\pi}$ and 4-momentum
          transfer values.  Dash-dotted lines show   
          kinematical limits. \label{fig:g11}}
\end{figure}

Results for the forward $C^{\pm}(\omega)$, forward derivative
$E^{\pm}(\omega)$, and fixed-t $C^{\pm}(\nu,t)$, H\"uper, and
$B^+(\nu,t)$ DRs are summarized graphically in 
Figures~\ref{fig:g2} $-$ \ref{fig:g4}, \ref{fig:g10}, and~
\ref{fig:g11}, respectively.  The form of $B^+(\nu,t)$ DR displayed
in Fig.~\ref{fig:g4} was used in Ref.~\cite{bugg_g2} in a determination
of the coupling constant. This form was not
used as a constraint in FA02, but is included here for its utility in
illustrating the uncertainty associated with $g^{2}/4\pi$.
These DRs are well satisfied over
the whole constraint region up to a kinetic energy T$_{\pi}$ = 1~GeV and a
four-momentum transfer $t = -0.4~(GeV/c)^2$. The consistency of
FA02 with these constraints is much better than 
our previous solution SM95~\cite{sm95} and the solutions
KA84~\cite{ka84} and KH80~\cite{kh80}. Due to the energy-dependent
parameterization of our partial waves, all fixed-t DRs
are well satisfied up to $t = -0.5~(GeV/c)^2$, with all
isovector DRs reasonably well satisfied up to
$\sim$2~GeV \textit{i.e.} the entire data range. The more sensitive
isoscalar DRs remain reasonably well satisfied up to
1100$-$1300~MeV; constraints for the forward $C^{+}(\omega)$ DR extended
to 0.8~GeV.

\begin{figure}[th]
\centering{
\includegraphics[height=0.4\textwidth, angle=90]{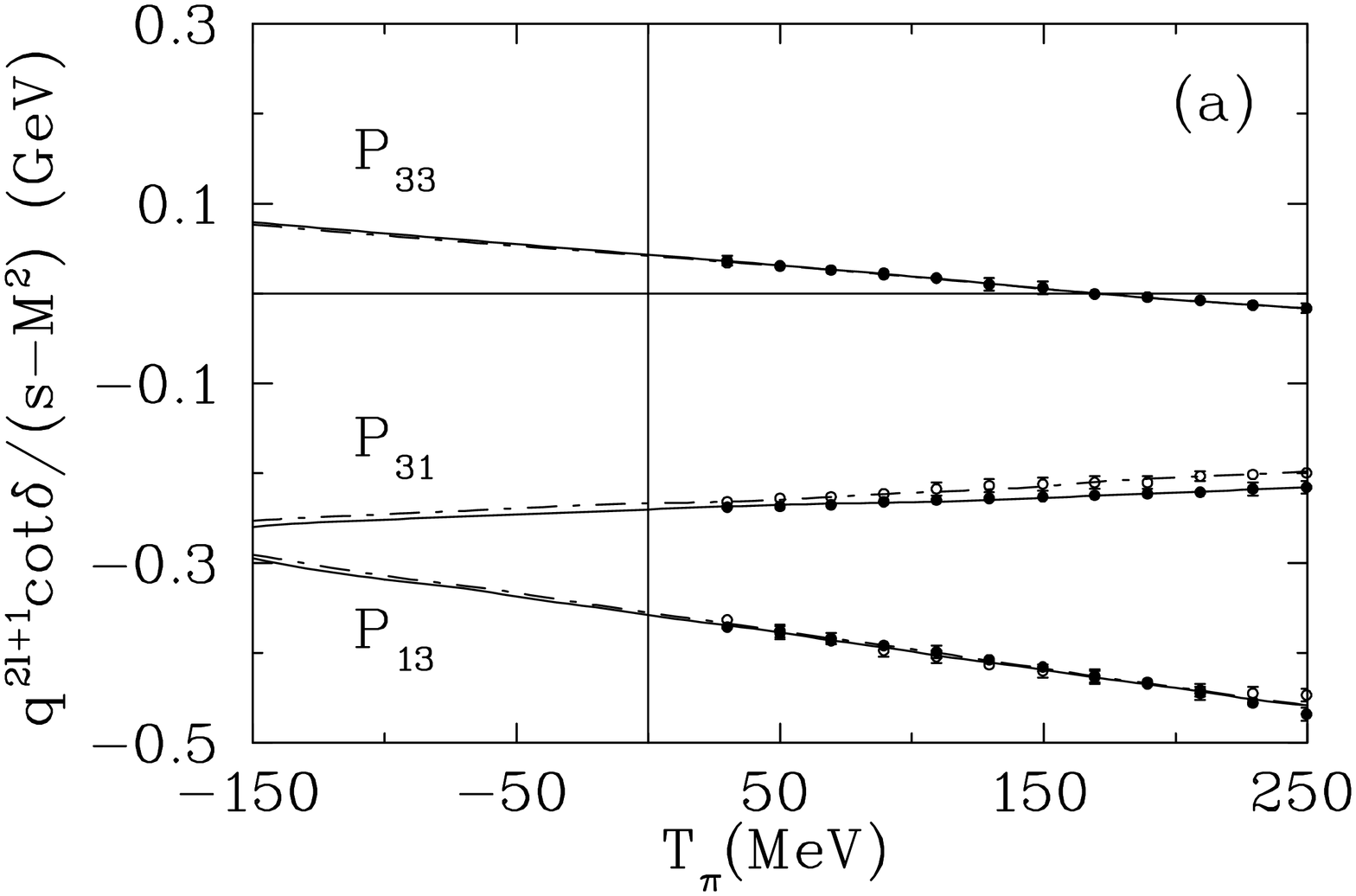}\hfill
\includegraphics[height=0.4\textwidth, angle=90]{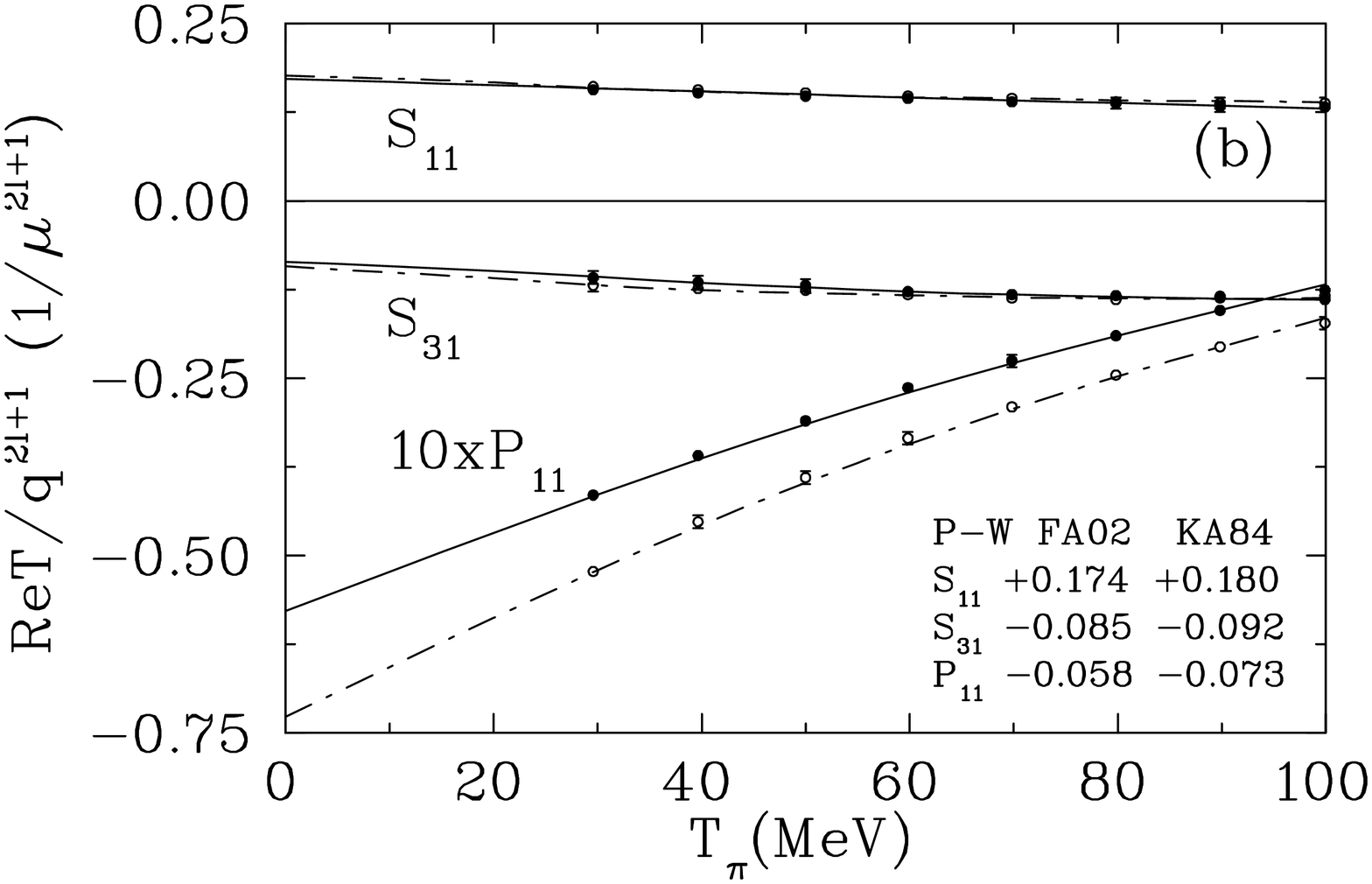} 
}\caption{Plots of (a) $q^{2l+1} \cot\delta/(s - M^2)$ and
          (b) Re$T$/$q^{2l+1}$ for FA02 (solid) and KA84~ 
          \protect\cite{ka84} (dot-dashed) S- and P-wave
          amplitudes.  Scattering lengths (volume) in 
          right panel given in $\mu^{-1}$ ($\mu^{-3}$) 
          units.  Curves are fits quadratic in $T_{\pi}$. 
          \label{fig:g12}}
\end{figure}

Figure~\ref{fig:g12} compares the ``Chew-Low"~\cite{chew-low} and 
scattering length (volume) plots for the FA02 and Karlsruhe~\cite{ka84} 
S- and P-wave amplitudes.  Exact agreement with the values given in 
Fig.\ref{fig:g2} is not expected, as the method is different. The FA02 
Chew-Low plots have a very similar behavior to the partial-wave dispersion
relation (PWDR) constrained KA84 solution. This indicates that FA02 at
least approximately satisfies a P-wave PWDR.  The present solution fixes
a long-standing discrepancy in the $P_{13}$ Chew-Low behavior
observed in our previous solutions~\cite{fa93,sm90,sm95}.

The present solution FA02 yields the DR parameters
g$^{2}/4\pi$ = 13.75$\pm$0.10 for the pion-nucleon 
coupling constant, a$^+_{1+}$ = 0.133$\pm$0.001~
$\mu^{-3}$ (a$^-_{1+}$ = -0.074$\pm$0.001~$\mu^{-3}$) 
for the P-wave scattering volumes, and 3a$^-_{0+}$ = 
0.2650$\pm$0.0014~$\mu^{-1}$ and a$_{\pi^-p}$ = 
0.0870$\pm$0.0013~$\mu^{-1}$ (implying a$^+_{0+}$ = 
-0.0010$\pm$0.0012~$\mu^{-1}$) for the S-wave 
scattering lengths.  The coupling constant confirms our 
earlier result in Ref.~\cite{sm95} and is in line with  
other determinations~\cite{deSw97,sc01}. The scattering 
lengths are in agreement with those of the PSI 
pionic-hydrogen and pionic-deuterium experiments~
\cite{piat}.

Our DR procedure constrains the
$\pi^-$p scattering length to \textit{agree} (within some error bound)
with the result from the PSI pionic-atom experiment~\cite{piat}
(a$_{\pi^-p}$ = 0.0883$\pm$0.0008~$\mu^{-1}$), so the above agreement in
a$_{\pi^-p}$ is not surprising.  However, a solution 
constructed to be otherwise identical to FA02, but where a$_{\pi^-p}$ was
\textit{not} constrained to the PSI value, yielded a
best fit with a$_{\pi^-p}$ = 0.0856$\pm$0.0010~$\mu^{-1}$, indicating
the value sought by the scattering data alone. Consequently, a
compromise value of $\sim$0.087 a little lower than the PSI result was
chosen as the constraint value for FA02. Ericson,
Loiseau, and Wycech have published~\cite{eric03} a reanalysis of the
PSI result~\cite{piat}, and have obtained a$^-_{0+}$ = 0.0883$\pm$0.0027~
$\mu^{-1}$ and a$_{\pi^-p}$ = 0.0870$\pm$0.0005~$\mu^{-1}$, 
the latter reasonably consistent with our result 0.0856$\pm$0.0010~
$\mu^{-1}$ using data alone.

The systematic check described above is one of many
that we have applied to our solution.  A large number of
test solutions were obtained with varying fitting and constraint
conditions, to investigate systematic uncertainties in the extracted
DR parameters. These systematics checks are discussed
in detail below. 

\begin{itemize}
\item
A solution (P370) using the point-Coulomb corrections
employed in our previous solutions Refs.~\cite{fa93,sm90,sm95} gave an
overall $\chi^2$ fit very similar to FA02.  
Differences in the resulting parameters were not large. For 
$g^{2}/4\pi$ and a$^-_{0+}$, this solution gave 13.69 and 
0.0867~$\mu^{-1}$, respectively. Despite seemingly large
differences between the point-Coulomb and
Nordita+Gibbs~\cite{trom,gibbsCoul} schemes (see
Fig.~\ref{fig:g5}), the consistency of results indicates that
our Coulomb correction scheme is adequate to
perform this type of partial-wave analysis.

\item
In an isospin-invariant framework, charge-exchange
data are not required to reconstruct the amplitudes.  As discussed
in Sec.~\ref{sec:csv}, a solution was constructed (X370) after 
charge-exchange data were removed from the fitted database. Despite
this drastic change, the resulting DR
parameters were quite consistent with those of FA02. 

\item
A less drastic change in the database was investigated in the region
of the $\Delta(1232)$ resonance. The issue of discrepant data sets on
the rising part of the resonance was discussed in Sec.~\ref{sec:expt}.  To
quantify this effect, solutions were constructed
employing the data sets of Pedroni \textit{et al.}~\cite{ped78}, Brack
\textit{et al.}~\cite{br86,br95}, and Pavan \textit{et al.}~\cite{pa01}, 
while floating
(applying a normalization error of 100\%) those of Bussey \textit{et al.}
~\cite{bu73} and Carter \textit{et al.}~\cite{ca71}, and vice versa.  The
$\pi^-$p scattering-length constraint was held constant in both cases.
The resulting solutions differed in the coupling constant by 
$\Delta g^{2}/4\pi$ = 0.07, 
while the scattering lengths remained consistent. Since
FA02 fits all the aforementioned data sets together, an uncertainty of
$\pm$0.04 was ascribed to this systematic effect.

\item
The DR constraints for the FA02 solution were applied
up to 1~GeV, and from $t$ = 0  to $-0.4~(GeV/c)^2$. Solutions
constructed with a reduced range of 25 to 600~MeV and $t$ to $-0.3~
(GeV/c)^2$ again yielded results consistent with FA02, the
latter ranges having been employed in solution SM95~\cite{sm95}.

\item
Dispersion relation constraints were both strengthened and weakened
by a factor of 2, yielding results consistent with FA02. It should 
be noted, however, that constraining too tightly yields a very large 
data $\chi^2$ and numerically unstable fits.

\item
In addition to the above test, a solution was constructed employing 
\textit{no} DR constraints. Here, the fit to data was naturally much 
better than in FA02, but the improvement was
less dramatic ($\Delta\chi^2$ = 1036 for 21808 data points) than one might
expect, with little difference obvious in a plot of the partial waves. The
numerically sensitive C$^+$ and E$^+$ isoscalar dispersion relations were
understandably not well satisfied; however, the less sensitive  C$^-$ and 
E$^-$ isovector dispersion relations remained rather well satisfied over
the full energy range, as were the H\"uper and B$^+$($\nu$,t) relations
up to $\sim$800~MeV and $t = -0.3~(GeV/c)^2$. The extracted constants 
were a$^-_{0+}$ = 0.086~$\mu^{-1}$,  a$^-_{1+}$ = $-$0.073~$\mu^{-3}$, and
g$^{2}/4\pi$ = 13.62.  

\item
The D-waves below $\sim$250~MeV are too small to be accurately 
determined from fits to data alone.  The forward derivative E$^{\pm}$
DRs have an enhanced sensitivity to higher partial
waves ($\sim l^3$) and so help to constrain the D-waves. To
test the dependence of the DR
parameters on these low energy D-waves, a
solution was constructed identical to FA02 except constrained
rigorously (1\%, compared to 5\% for FA02) to the low energy Koch
D-wave results~\cite{koch}. The DR parameters changed minimally, 
the largest changes being 0.002 for both
$\Delta$a$^+_{1+}$, and $\Delta$a$^{0}_{1+}$. However,
the data $\chi^2$  was increased both
above 400~MeV ($\Delta\chi^2$ = $\sim$600) and below by $\sim$200,
with an increase of $\sim$300 in the fit to DR pseudo-data
constraints.  The fit thus exhibits a clear preference for the FA02
D-waves. 

\end{itemize}

The central values and uncertainties of the DR
parameters have been estimated, taking into
consideration the systematic checks outlined above, and from other
checks within the FA02 solution itself.  These include differences in
parameter values at the $\chi^2$ minima for each of the charge
channels and the DR pseudo-data. For the coupling constant $g^2/4\pi$,
we also considered differences in the value extracted
from the H\"uper (Fig.~\ref{fig:g10}) and $B^+(\nu,t)$
(Fig.~\ref{fig:g11}) dispersion relations, which, along 
with the database changes around the $\Delta(1232)$, 
were the dominant sources of uncertainty.  The 
statistical uncertainty stemming from $\Delta\chi^2 = 
1$ variations from the minima is negliglible compared to 
systematic effects.

These tests for systematic effects improve our 
confidence in the values of parameters extracted from FA02. It
should be noted, however, that the scattering lengths are tied to the
PSI pionic-atom value for the $\pi^-$p scattering length.  A new
more precise experiment is planned~\cite{piat2}, and if the result
were to change several standard deviations from the present value, this would
necessitate a re-analysis. Nonetheless, the value of the  pion-nucleon coupling
constant would not be expected to change, as it is observed to be robust
with respect to modest changes in the $\pi^-$p scattering length.

\subsection{The $\pi^-p\to\eta n$ Channel}
\label{sec:eta}

\begin{figure}[th]
\centering{
\includegraphics[height=0.7\textwidth, angle=90]{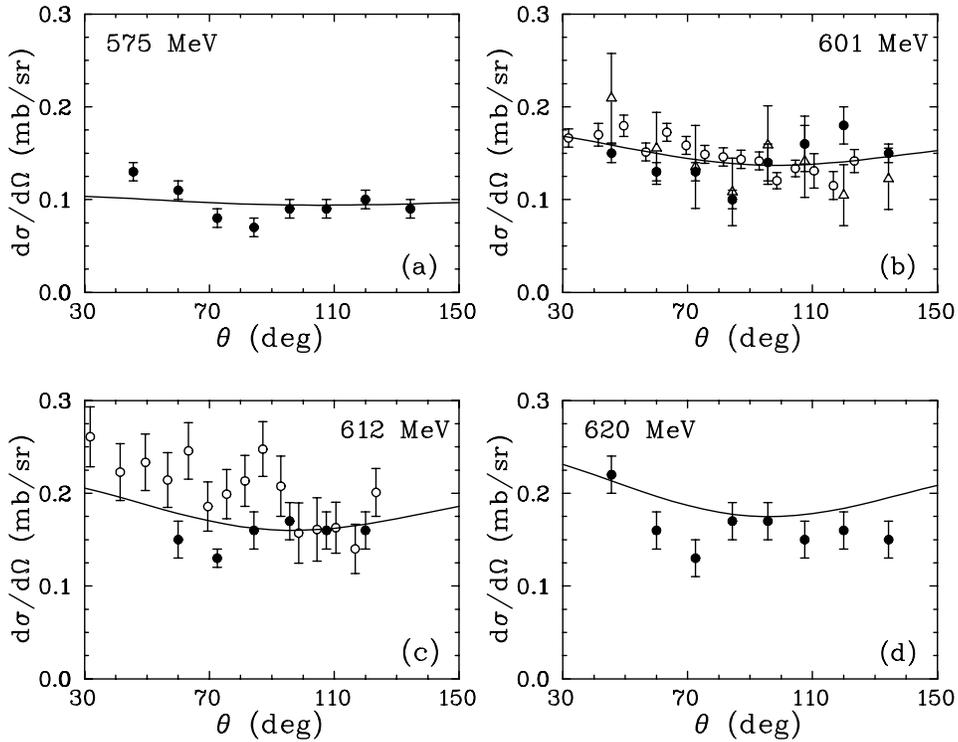}
}\caption{Differential cross sections for $\pi^-p\to\eta n$.
          (a) 575~MeV, (b) 601~MeV, (c) 612~MeV, and (d)
          620~MeV.  Experimental data are from
          ~\protect\cite{e909} measurements (filled circles),
          ~\protect\cite{de69} (open circles), and   
          ~\protect\cite{ri70} (open triangles).
          Solid line shows the FA02 results.
          \label{fig:g13}}
\end{figure}

Our fit to a representative set of $\pi^-p\to\eta n$ 
cross sections is displayed in Fig.~\ref{fig:g13}.  In 
general, over this region, our results are 
qualitatively similar to other recent multi-channel 
analyses which have included additional reaction 
data~\cite{multi}.  Compared to the $\pi$N elastic 
result, the overall $\chi^2$ for this channel is 
slightly higher.  However, we feel this is largely due 
to problems in the database.

\begin{figure}[th]
\centering{
\includegraphics[height=0.7\textwidth, angle=90]{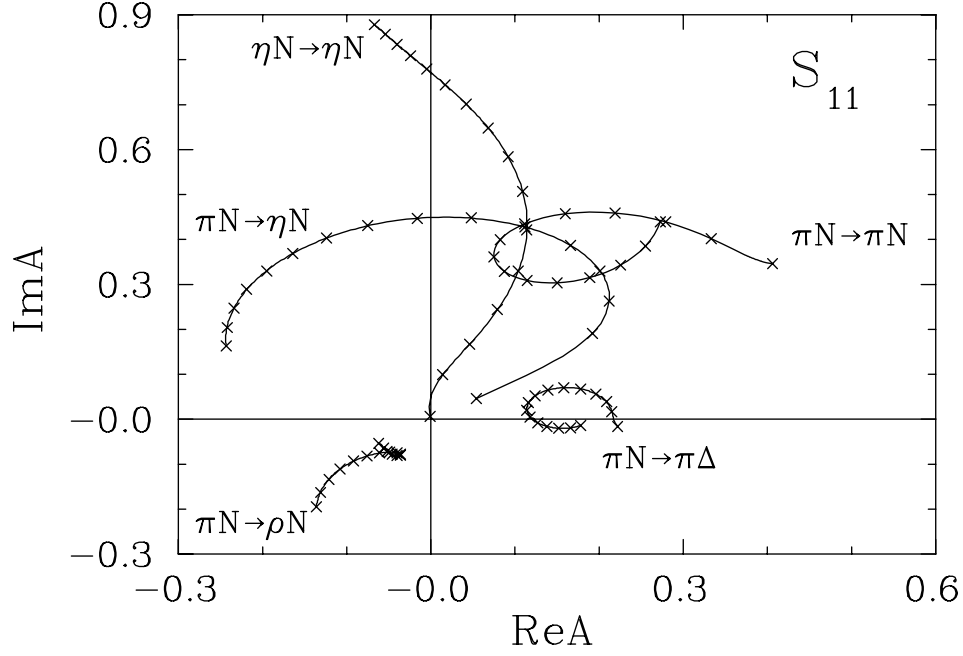}
}\caption{Essential S$_{11}$ amplitudes from threshold to
          800~MeV (W = 1487 to 1623~MeV).  Crosses
          indicate every 10~MeV step in W. \label{fig:g14}}
\end{figure}

\begin{figure}[th]
\centering{
\includegraphics[height=0.4\textwidth, angle=90]{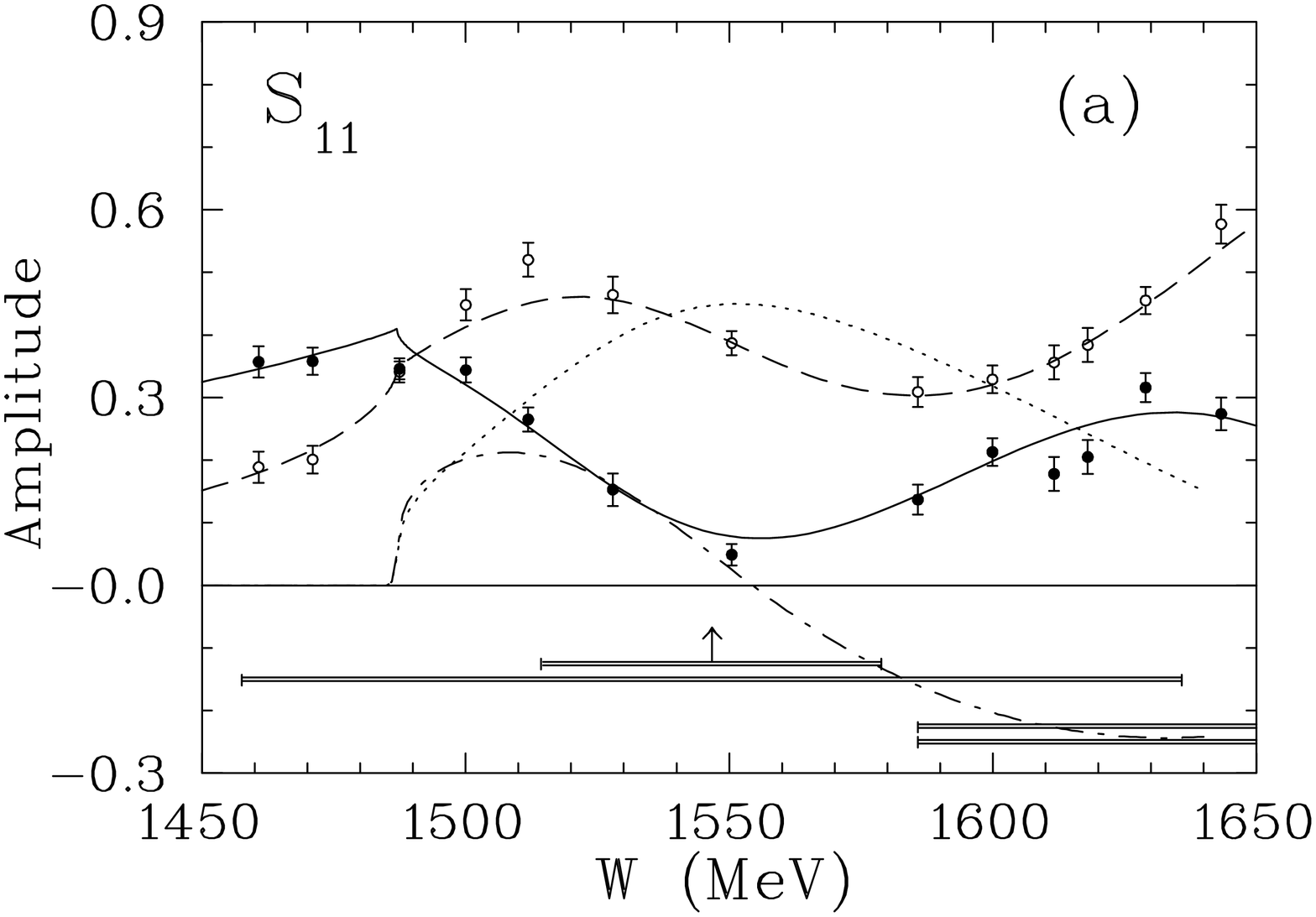}\hfill
\includegraphics[height=0.4\textwidth, angle=90]{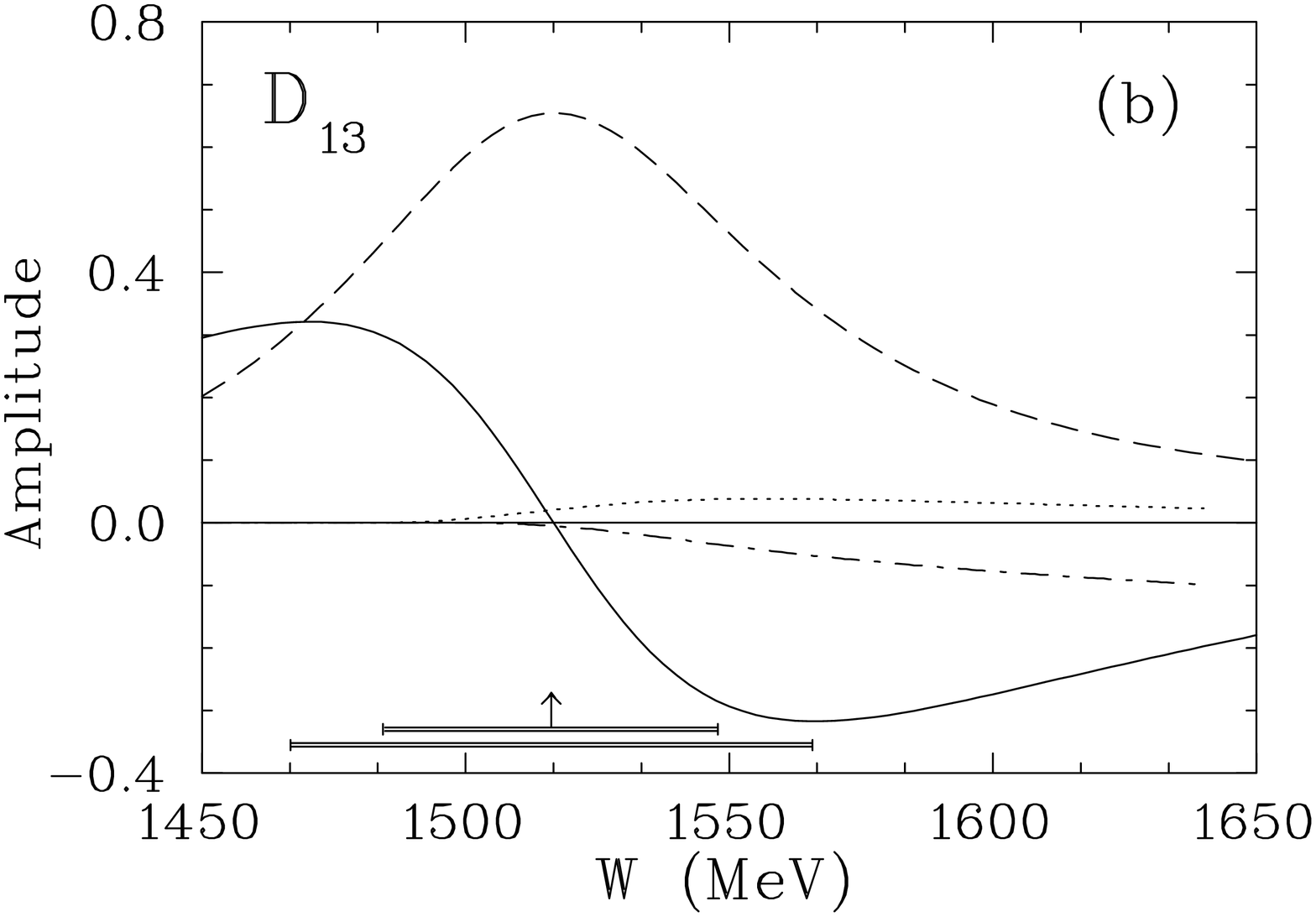}
}\caption{Low-lying states related to the $\eta N$ interaction.
      Plotted are the results for (a) $S_{11}$ and (b)
      $D_{13}$ amplitudes.  Solid (dashed) curves give the
      real (imaginary) parts of $\pi N\to\pi N$ amplitudes,
      dash-dotted (dotted) curves represent the real
      (imaginary) parts of $\pi N\to\eta N$ amplitudes
      corresponding to the FA02 solution.  All amplitudes
      are dimensionless as in Fig.~\protect\ref{fig:g6}.
      Vertical arrows indicate $W_R$ and horizontal bars
      show full $\Gamma$/2 and partial widths for $\Gamma
      _{\pi N}$ associated with the FA02 results.
      \label{fig:g15}}
\end{figure}

Our coupled-channel approach allows the determination 
of a number of amplitudes related to the $\eta$N 
interaction.  Fig.~\ref{fig:g14} gives an 
Argand~\cite{argand} plot representation of the 
energy-dependent fit FA02 over the energy region including 
$\pi^-p\to\eta n$ data. Fig.~\ref{fig:g15} gives a more 
detailed view of our $\pi$N elastic and $\pi N\to\eta 
N$ results for the $S_{11}$ and $D_{13}$ partial waves. 

\section{Summary and Conclusion}
\label{sec:conc}

We have fitted the existing $\pi$N elastic scattering database, 
employing a complete set of DR
constraints, up to T$_{\pi}$ = 1~GeV and $t = -0.4~(GeV/c)^2$.
Data from the reaction $\pi^-p\to\eta n$ have been included for 
the first time in an 
analysis of this type. These improvements have allowed us to 
more carefully examine the N(1535), which is nearly obscured 
(in the $\pi$N elastic reaction) by the opening of the $\eta$N 
threshold.  Remarkably, the resulting S$_{11}$ partial wave 
shows little change over this energy region.  However, the 
N(1535) resonance width has changed (increased) dramatically, 
this being due mainly to our new method for fitting Breit-Wigner 
parameters. This result should be taken into account by any 
multi-channel analysis which fits single-energy partial-wave 
amplitudes. 

Our fits without DR 
constraints and without charge-exchange
data have also yielded interesting results. In most cases,
an extensive use of DRs had little effect on the extracted
resonance spectrum. For weak or non-canonical structures,
however, dispersion relations may play a more important role. 

In the absence
of charge symmetry violation (CSV),
beyond Coulomb effects, one should be able to predict
the charge-exchange observables based on a fit to elastic
scattering data. In the low-energy region, where CSV effects
are expected to be most important, our fit excluding 
charge-exchange data is reasonably predictive, and could be 
useful in a more refined study of this issue.

An extensive list of tests designed to check for systematic effects
(changes to the database, Coulomb-correction scheme) has 
(a) revealed the DR parameters to be
fairly robust with respect to these
effects, and (b) suggested a way to quantify these systematic
uncertainties. As the FA02 the solution exhibits
good consistency with the complete set of forward and fixed-t
DRs, we have added confidence in our results. In
particular, the $\pi$NN coupling has remained quite stable, changing
little from our SM95 determination.  Our DR analysis and
systematic checks will next be applied to an
extraction of the nucleon $\sigma$-term (see,
\textit{e.g.} Ref.~\cite{glls88}).  The extraction of this quantity is
sensitive to the fine details of the analysis, and a careful
examination of the $\sigma$-term is in preparation.

\acknowledgments

The authors express their gratitude to B.~Bassalleck, V.~S.~
Bekrenev, J.~T.~Brack, M.~Clajus, H.~Crannel, 
J.~Comfort, E.~Friedman, G.~J.~Hofman, C.~V.~Gaulard, L.~D.~
Isenhover, M.~Janousch, G.~Jones, V.~P.~Kanavets, M.~A.~
Kovash, N.~Kozlenko, A.~A.~Kulbardis, I.~V.~Lopatin, M.~
Mikuz, R.~Meier, V.~E.~Markushin, T.~W.~Morrison, D.~
Po\v{c}ani\'{c}, R.~A.~Ristinen, M.~E.~Sadler, G.~R.~Smith, 
V.~V.~Sumachev, P.~Weber, and R.~Wieser for providing 
experimental data prior to publication or for clarification 
of information already published.  We are grateful to 
W.~R.~Gibbs for providing us with a calculation of 
finite-source-size Coulomb barrier corrections.  We also 
acknowledge useful communications over the years with G.~
H\"ohler, D.~V.~Bugg, and D.~M.~Manley.  This work was 
supported in part by the U.~S.~Department of Energy under 
Grant DE--FG02--99ER41110.  The authors (R.~A., I.~S., and 
R.~W.) acknowledge partial support from Jefferson Lab and 
the Southeastern Universities Research Association under 
DOE contract DE--AC05--84ER40150.


\newpage
\begin{table}[th]
\caption{Comparison of present (FA02) and previous
         (SM95~\protect\cite{sm95}, FA93~
         \protect\cite{fa93}, SM90~\protect\cite{sm90},
         and FA84~\protect\cite{fa84}) energy-dependent 
         partial-wave analyses of elastic $\pi^\pm p$, 
         charge-exchange ($\pi^0n$), and $\pi^-p\to\eta 
         n$ ($\eta n$) scattering data.  For FA02 
         solution, $\eta N$ data has been included to 
         800~MeV.  The $\chi^2$ values for the previous 
         solutions correspond to our published results.  
         $N_{prm}$ is the number of parameters ($I = 1/2$ 
         and $3/2$) varied in the fit. \label{tbl1}}
\begin{tabular}{ccccccc}
\colrule
Solution & Range~(MeV) &
$\chi^2$/$\pi^+p$ & $\chi^2$/$\pi^-p$ & $\chi^2$/$\pi^0n$ &
$\chi^2$/$\eta n$ & $N_{prm}$  \\
\colrule
FA02 & 2100 & 21735/10468 & 18932/9650 & 4136/1690 & 439/173 & 94/78 \\
SM95 & 2100 & 23593/10197 & 18855/9421 & 4442/1625 &   $-$   & 94/80 \\
FA93 & 2100 & 23552/10106 & 20747/9304 & 4834/1668 &   $-$   & 83/77 \\
SM90 & 2100 & 24897/10031 & 24293/9344 &10814/2132 &   $-$   & 76/68 \\
FA84 & 1100 &  7416/ 3771 & 10658/4942 & 2062/ 717 &   $-$   & 64/57 \\
\colrule
\end{tabular}
\end{table}
\begin{table}[th]
\caption{Comparison of $\chi^2$/data for normalized (Norm)
         and unnormalized (Unnorm) data used in the FA02 solution.
         \label{tbl2}}
\begin{tabular}{ccc}
\colrule
Reaction & Norm & Unnorm \\
\colrule
$\pi^+p\to\pi^+p$ & 2.1 & 9.3 \\
$\pi^-p\to\pi^-p$ & 2.0 & 7.1 \\
$\pi^-p\to\pi^0n$ & 2.4 & 9.5 \\
$\pi^0n\to\eta n$ & 2.5 & 4.6 \\
\colrule
\end{tabular}
\end{table}
\newpage
\begin{table}[th]
\caption{Single-energy (binned) fits of combined elastic 
         $\pi^\pm p$, charge-exchange, $\pi^-p\to\eta n$ 
         scattering data, and $\chi^2$ values.  $N_{prm}$ 
         is the number of parameters varied in the 
         single-energy fits, and $\chi^2_E$ is given by  
         the energy-dependent fit, FA02, over the same 
         energy interval. \label{tbl3}}
\begin{tabular}{ccccc|ccccc}
\colrule
T$_{\pi}$~(MeV)&Range~(MeV)&$N_{prm}$&$\chi^2$/data&$\chi^2_E$&
T$_{\pi}$~(MeV)&Range~(MeV)&$N_{prm}$&$\chi^2$/data&$\chi^2_E$ \\
\colrule
  30 &  26 $-$  33 &  4 & 171/141 & 218 &  795 & 793 $-$ 796 & 30 & 204/165 & 301 \\
  47 &  45 $-$  49 &  4 &  74/ 82 &  80 &  820 & 813 $-$ 827 & 30 & 399/302 & 436 \\
  66 &  61 $-$  69 &  4 & 183/136 & 198 &  868 & 864 $-$ 870 & 31 & 294/211 & 389 \\
  90 &  87 $-$  92 &  4 & 116/111 & 154 &  888 & 886 $-$ 890 & 32 & 174/144 & 291 \\
 112 & 107 $-$ 117 &  6 & 114/ 93 & 115 &  902 & 899 $-$ 905 & 34 & 539/416 & 891 \\
 124 & 121 $-$ 126 &  6 &  84/ 60 &  92 &  927 & 923 $-$ 930 & 34 & 234/200 & 378 \\
 142 & 139 $-$ 146 &  6 & 230/159 & 213 &  962 & 953 $-$ 971 & 34 & 387/299 & 593 \\
 170 & 165 $-$ 174 &  6 & 177/141 & 170 & 1000 & 989 $-$1015 & 36 & 691/423 & 847 \\
 193 & 191 $-$ 195 &  6 & 103/107 & 119 & 1030 &1022 $-$1039 & 38 & 286/272 & 396 \\
 217 & 214 $-$ 220 &  6 & 116/109 & 143 & 1044 &1039 $-$1048 & 38 & 362/243 & 520 \\
 238 & 235 $-$ 241 &  6 & 124/115 & 149 & 1076 &1073 $-$1078 & 38 & 240/218 & 432 \\
 266 & 263 $-$ 271 &  6 & 162/123 & 175 & 1102 &1099 $-$1103 & 39 & 232/173 & 346 \\
 292 & 291 $-$ 293 &  8 & 157/129 & 199 & 1149 &1147 $-$1150 & 40 & 339/210 & 468 \\
 309 & 306 $-$ 310 &  8 & 160/140 & 174 & 1178 &1165 $-$1192 & 41 & 759/394 & 925 \\
 334 & 332 $-$ 335 &  8 &  94/ 58 & 130 & 1210 &1203 $-$1216 & 42 & 286/233 & 364 \\
 352 & 351 $-$ 352 &  9 &  82/110 &  99 & 1243 &1237 $-$1248 & 44 & 473/283 & 586 \\
 389 & 387 $-$ 390 & 10 &  31/ 28 &  74 & 1321 &1304 $-$1337 & 44 & 731/401 & 946 \\
 425 & 424 $-$ 425 & 10 & 151/139 & 191 & 1373 &1371 $-$1374 & 46 & 331/166 & 613 \\
 465 & 462 $-$ 467 & 13 & 358/120 & 451 & 1403 &1389 $-$1417 & 48 & 544/408 & 811 \\
 500 & 499 $-$ 501 & 16 & 161/136 & 209 & 1458 &1455 $-$1460 & 48 & 279/258 & 504 \\
 518 & 515 $-$ 520 & 18 & 104/ 78 & 144 & 1476 &1466 $-$1486 & 48 & 510/344 & 682 \\
 534 & 531 $-$ 535 & 18 & 131/128 & 184 & 1570 &1554 $-$1586 & 48 & 834/546 &1062 \\
 560 & 557 $-$ 561 & 19 & 310/151 & 582 & 1591 &1575 $-$1606 & 48 & 402/336 & 588 \\
 580 & 572 $-$ 590 & 20 & 382/286 & 556 & 1660 &1645 $-$1673 & 48 & 510/391 & 732 \\
 599 & 597 $-$ 600 & 22 & 253/151 & 382 & 1720 &1705 $-$1734 & 48 & 388/279 & 510 \\
 625 & 622 $-$ 628 & 24 & 125/ 94 & 204 & 1753 &1739 $-$1766 & 48 & 654/439 & 864 \\
 662 & 648 $-$ 675 & 25 & 554/352 & 735 & 1838 &1829 $-$1845 & 48 & 444/290 & 698 \\
 721 & 717 $-$ 725 & 28 & 160/129 & 217 & 1875 &1852 $-$1897 & 48 & 948/674 &1284 \\
 745 & 743 $-$ 746 & 28 & 159/100 & 286 & 1929 &1914 $-$1942 & 48 & 857/501 &1187 \\
 765 & 762 $-$ 767 & 30 & 190/167 & 289 & 1970 &1962 $-$1978 & 48 & 461/271 & 684 \\
 776 & 774 $-$ 778 & 30 & 220/155 & 302 & 2026 &2014 $-$2037 & 48 & 368/320 & 652 \\
\colrule
\end{tabular}
\end{table}
\newpage
\begin{table}[th]
\caption{Resonance couplings from a Breit-Wigner fit to
         the FA02 solution {[}GW{]}, our previous solution
         SM95 {[}VPI{]}~\protect\cite{sm95}, and an average
         from the Review of Particle Properties {[}RPP{]}~
         \protect\cite{rpp} (in square brackets).  Masses
         W$_R$, half-widths $\Gamma$/2, and partial widths
         for $\Gamma_{\pi N}$/$\Gamma$ are listed for isospin
         $1/2$ baryon resonances $N^{\ast}$. \label{tbl4}}
\begin{tabular}{ccccc}
\colrule
Resonance      & W$_R$ & $\Gamma$/2  & $\Gamma_{\pi N}/ \Gamma$ & Ref\\
               & (MeV) & (MeV)       &                          &    \\
\colrule
$P_{11}(1440)$ & 1468.0$\pm$4.5&180$\pm$13&0.750$\pm$0.024      & GW \\
               & 1467  &  220        &  0.68                    & VPI\\
               &[1440] & [175]       & [0.65]                   & RPP\\
$D_{13}(1520)$ & 1516.3$\pm$0.8&49.3$\pm$1.3&0.640$\pm$0.005    & GW \\
               & 1515  &   53        &  0.61                    & VPI\\
               &[1520] &  [60]       & [0.55]                   & RPP\\
$S_{11}(1535)$ & 1546.7$\pm$2.2&89.0$\pm$5.8&0.360$\pm$0.009    & GW \\
               & 1535  &   33        &  0.31                    & VPI\\
               &[1535] &  [75]       & [0.45]                   & RPP\\
$S_{11}(1650)$ & 1651.2$\pm$4.7&65.3$\pm$3.5&1.000              & GW \\
               & 1667  &   45        & $\approx$1.0             & VPI\\
               &[1650] &  [75]       & [0.72]                   & RPP\\
$D_{15}(1675)$ & 1676.2$\pm$0.6&75.9$\pm$1.5&0.400$\pm$0.002    & GW \\
               & 1673  &   77        &  0.38                    & VPI\\
               &[1675] &  [75]       & [0.45]                   & RPP\\
$F_{15}(1680)$ & 1683.2$\pm$0.7&67.2$\pm$1.9&0.670$\pm$0.004    & GW \\
               & 1678  &   63        &  0.68                    & VPI\\
               &[1680] &  [65]       & [0.65]                   & RPP\\
$P_{13}(1720)$ & 1749.6$\pm$4.5&128$\pm$11&0.190$\pm$0.004      & GW \\
               & 1820  &  177        &  0.16                    & VPI\\
               &[1720] &  [75]       & [0.15]                   & RPP\\
$G_{17}(2190)$ & 2192.1$\pm$8.7&363$\pm$31&0.230$\pm$0.002      & GW \\
               & 2131  &  238        &  0.23                    & VPI\\
               &[2190] & [225]       & [0.15]                   & RPP\\
$H_{19}(2220)$ & 2270$\pm$11&183$\pm$21&0.200$\pm$0.006         & GW \\
               & 2258  &  167        &  0.26                    & VPI\\
               &[2220] & [200]       & [0.15]                   & RPP\\
$G_{19}(2250)$ & 2376$\pm$43&462$\pm$89&0.110$\pm$0.004         & GW \\
               & 2291  &  386        &  0.10                    & VPI\\
               &[2250] & [200]       & [0.10]                   & RPP\\
\colrule
\end{tabular}
\end{table}
\newpage
\begin{table}[th]
\caption{Parameters for isospin $3/2$ baryon resonances
         $\Delta^{\ast}$.  Notation as in Table~
         \protect\ref{tbl4}. \label{tbl5}}
\begin{tabular}{ccccc}
\colrule
Resonance      & W$_R$ & $\Gamma$/2  & $\Gamma_{\pi N}/ \Gamma$ & Ref\\
               & (MeV) & (MeV)       &                          &    \\
\colrule
$P_{33}(1232)$ & 1232.9$\pm$1.2&59.0$\pm$1.1&1.000              & GW \\
               & 1233  &   57        & $\approx$1.0             & VPI\\
               &[1232] &  [60]       & [$>$0.99]                & RPP\\
$S_{31}(1620)$ & 1614.1$\pm$1.1&70.5$\pm$3.0&0.310$\pm$0.004    & GW \\
               & 1617  &   54        &  0.29                    & VPI\\
               &[1620] &  [75]       & [0.25]                   & RPP\\
$D_{33}(1700)$ & 1687.9$\pm$2.5&182.4$\pm$8.3&0.150$\pm$0.001   & GW \\
               & 1680  &  136        &  0.16                    & VPI\\
               &[1700] & [150]       & [0.15]                   & RPP\\
$F_{35}(1905)$ & 1855.7$\pm$4.2&167$\pm$11&0.120$\pm$0.002      & GW \\
               & 1850  &  147        &  0.12                    & VPI\\
               &[1905] & [175]       & [0.10]                   & RPP\\
$D_{35}(1930)$ & 2046$\pm$45&201$\pm$99&0.040$\pm$0.014         & GW \\
               & 2056  &  295        &  0.11                    & VPI\\
               &[1930] & [175]       & [0.15]                   & RPP\\
$F_{37}(1950)$ & 1923.3$\pm$0.5&139.1$\pm$1.5&0.480$\pm$0.002   & GW \\
               & 1921  &  116        &  0.49                    & VPI\\
               &[1950] & [150]       & [0.37]                   & RPP\\
\colrule
\end{tabular}
\end{table}
\newpage
\begin{table}[th]
\caption{Pole positions from the
         solution FA02 {[}GW{]}, our previous
         solution SM95 {[}VPI{]}~\protect\cite{sm95}, and
         an average from the Particle Data Group {[}RPP{]}~
         \protect\cite{rpp} (in square brackets).  $Re$
         ($W_R$) and $Im$ ($W_I$) parts are listed for
         isospin $1/2$ baryon resonances.  Second 
         sheet poles are labeled by $\dagger$.  Modulus 
         and phase 
         values are listed for the $\pi N$ elastic pole 
         residue. \label{tbl6}}
\begin{tabular}{cccccccccccc}
\colrule
Wave     & $W_R$ &$-W_I$ &Modulus& Phase & Ref & Wave     & $W_R$ &$-W_I$&Modulus& Phase & Ref\\
         & (MeV) & (MeV) & (MeV) & (deg) &     &          & (MeV) & (MeV)& (MeV) & (deg) &    \\
\colrule
$S_{11}$ & 1526  &   65  &  33   &    14 & GW  & $D_{15}$ & 1659  &   73&  29   &  $-$22 & GW \\
         & 1501  &   62  &  31   & $-$12 & VPI &          & 1663  &   76&  29   &  $-$6  & VPI\\
         &[1505] &  [85] & $-$   & $-$   & RPP &          &[1660] &  [70]& $-$  & $-$    & RPP\\
$S_{11}$ & 1653  &   91  &  69   & $-$55 & GW  & $F_{15}$ & 1678  &   60 &  43   &     1 & GW \\
         & 1673  &   41  &  22   &    29 & VPI &          & 1670  &   60 &  40   &     1 & VPI\\
         &[1660] &  [80] & $-$   & $-$   & RPP &          &[1670] &  [60]& $-$   & $-$   & RPP\\
$P_{11}$ & 1357  &   80  &  36   &$-$102 & GW  & $F_{15}$ & 1779  &  124 &  47   & $-$61 & GW \\
         & 1346  &   88  &  42   &$-$101 & VPI &          & 1793  &   94 &  27   & $-$56 & VPI\\
         &[1365] & [105] & $-$   & $-$   & RPP &          &  $-$  &   $-$& $-$   & $-$   & RPP\\
$P_{11}^\dagger$
         & 1385  &   83  &  82   & $-$51 & GW  & $G_{17}$ & 2076  &  251 &  68   & $-$32 & GW \\
         & 1383  &  105  &  92   & $-$54 & VPI &          & 2030  &  230 &  46   & $-$23 & VPI\\
         &  $-$  &   $-$ & $-$   & $-$   & RPP &          &[2050] & [225]& $-$   & $-$   & RPP\\
$P_{13}$ & 1655  &  139  &  20   & $-$88 & GW  & $G_{19}$ & 2238  &  268 &  33   & $-$25 & GW \\ 
         & 1717  &  194  &  39   & $-$70 & VPI &          & 2087  &  340 &  24   & $-$44 & VPI\\
         &[1700] & [125] & $-$   & $-$   & RPP &          &[2140] & [240]& $-$   & $-$   & RPP\\
$D_{13}$ & 1514  &   51  &  35   & $-$6  & GW  & $H_{19}$ & 2209  &  282 &  96   & $-$71 & GW \\
         & 1515  &   55  &  34   &     7 & VPI &          & 2203  &  268 &  68   & $-$43 & VPI\\
         &[1510] &  [57] & $-$   & $-$   & RPP &          &[2170] & [235]& $-$   & $-$   & RPP\\
\colrule
\end{tabular}
\end{table}
\newpage
\begin{table}[th]
\caption{Pole positions for isospin $3/2$ baryon
         resonances $\Delta^{\ast}$.  Notation as in
         Table~\protect\ref{tbl6}. \label{tbl7}}
\begin{tabular}{cccccc}
\colrule
Wave     & $W_R$ &$-W_I$ &Modulus& Phase & Ref \\
         & (MeV) & (MeV) & (MeV) & (deg) &     \\
\colrule
$S_{31}$ & 1594  &   59  &  17   &$-$104 & GW \\
         & 1585  &   52  &  14   &$-$121 & VPI\\
         &[1600] &  [57] & $-$   & $-$   & RPP\\
$P_{31}$ & 1748  &  262  &  48   &   158 & GW \\
         & 1810  &  247  &  53   &$-$176 & VPI\\
         &[1855] & [175] & $-$   & $-$   & RPP\\
$P_{33}$ & 1210  &   50  &  53   & $-$47 & GW \\
         & 1211  &   50  &  38   & $-$22 & VPI\\
         &[1210] &  [50] & $-$   & $-$   & RPP\\
$D_{33}$ & 1617  &  113  &  16   & $-$47 & GW \\
         & 1655  &  121  &  16   & $-$12 & VPI\\
         &[1660] & [100] & $-$   & $-$   & RPP\\
$D_{35}$ & 1966  &  182  &  16   & $-$21 & GW \\
         & 1913  &  123  &   8   & $-$47 & VPI\\
         &[1890] & [125] & $-$   & $-$   & RPP\\
$F_{35}$ & 1825  &  135  &  16   & $-$25 & GW \\
         & 1832  &  127  &  12   & $-$4  & VPI\\
         &[1830] & [140] & $-$   & $-$   & RPP\\
$F_{37}$ & 1874  &  118  &  57   & $-$34 & GW \\
         & 1880  &  118  &  54   & $-$17 & VPI\\ 
         &[1885] & [120] & $-$   & $-$   & RPP\\ 
\colrule
\end{tabular}
\end{table}
\newpage
\begin{table}[th]
\caption{Zero positions from the
         solution FA02 {[}GW{]} and our previous
         solution SM95 {[}VPI{]}~\protect\cite{sm95}.  $Re$
         ($W_R$) and $Im$ ($W_I$) parts are listed for
         isospin $1/2$ and $3/2$ baryon resonances.
         \label{tbl8}}
\begin{tabular}{cccccccccccc}
\colrule
Wave     & $W_R$ &$-W_I$ & Ref \\
         & (MeV) & (MeV) &     \\
\colrule
$S_{11}$ & 1578  &   38  & GW \\
         & 1582  &   54  & VPI\\
         & 1695  &   43  & VPI\\
$P_{13}$ & 1585  &   51  & GW \\
         & 1618  &   85  & VPI\\
$D_{13}$ & 1759  &   64  & GW \\
         & 1751  &   87  & VPI\\
$F_{15}$ & 1765  &   66  & GW \\
         & 1775  &   57  & VPI\\
\colrule
$S_{31}$ & 1580  &   36  & GW \\
         & 1579  &   30  & VPI\\
$P_{31}$ & 1826  &  197  & GW \\
         & 1863  &  170  & VPI\\
$D_{35}$ & 1806  &  107  & GW \\
         & 1827  &   69  & VPI\\
\colrule
\end{tabular}
\end{table}  
\newpage
\begin{table}[th]
\caption{Comparison of FA02 and Breit-Wigner+background fits.
         Background parameters for isospin $1/2$ baryon
         resonance fits (see text and associated Table~
         \protect\ref{tbl4}).  ``Data" refers to the number 
         of scattering data used in the fit. \label{tbl9}}
\begin{tabular}{ccccccccc}
\colrule
Resonance      & Wmin & Wmax & BW fit   & FA02     & Data &   A              &       B         &  C  \\
               & (MeV)& (MeV)& $\chi^2$ & $\chi^2$ &      &                  &                 &     \\
\colrule
$P_{11}(1440)$ & 1350 & 1550 & 5556     & 5587     & 2393 & -0.270$\pm$0.026 &      $-$        & $-$ \\
$D_{13}(1520)$ & 1480 & 1560 & 3238     & 3341     & 1448 & -0.035$\pm$0.013 &      $-$        & $-$ \\
$S_{11}(1535)$ & 1490 & 1590 & 3566     & 3657     & 1632 &  0.342$\pm$0.018 & 4.530$\pm$0.851 & $-$ \\
$S_{11}(1650)$ & 1600 & 1720 & 6992     & 7218     & 3394 & -0.584$\pm$0.073 & 2.415$\pm$1.356 & 0.414$\pm$0.024 \\
$D_{15}(1675)$ & 1610 & 1730 & 7343     & 7359     & 3546 & -0.034$\pm$0.004 &      $-$        & $-$ \\
$F_{15}(1680)$ & 1620 & 1730 & 6683     & 6650     & 3172 &  0.011$\pm$0.011 & 1.192$\pm$0.575 & $-$ \\
$P_{13}(1720)$ & 1650 & 1790 & 8321     & 8484     & 4160 & -0.127$\pm$0.008 & 1.463$\pm$0.239 & $-$ \\
$G_{17}(2190)$ & 2049 & 2249 & 8017     & 8040     & 3774 &  0.024$\pm$0.009 &      $-$        & $-$ \\
$H_{19}(2220)$ & 2140 & 2250 & 5675     & 5682     & 2554 &  0.028$\pm$0.010 &      $-$        & $-$ \\
$G_{19}(2250)$ & 2010 & 2258 &11020     &11025     & 5330 & -0.046$\pm$0.010 &      $-$        & $-$ \\
\colrule
\end{tabular}
\end{table}
\begin{table}[th]
\caption{Comparison of FA02 and Breit-Wigner+background fits.
         Background parameters for isospin $3/2$ baryon
         resonance fits (see text and associated Table~
         \protect\ref{tbl5}). ``Data" refers to the number
         of scattering data used in the fit. \label{tbl10}}
\begin{tabular}{ccccccc}
\colrule
Resonance      & Wmin & Wmax & BW fit   & FA02     & Data &   A              \\
               & (MeV)& (MeV)& $\chi^2$ & $\chi^2$ &      &                  \\
\colrule
$P_{33}(1232)$ & 1180 & 1270 & 1180     & 1185     &  920 &  0.035$\pm$0.017 \\
$S_{31}(1620)$ & 1570 & 1680 & 5187     & 5212     & 2321 & -0.851$\pm$0.013 \\
$D_{33}(1700)$ & 1610 & 1770 & 9624     & 9690     & 4725 & -0.145$\pm$0.003 \\
$F_{35}(1905)$ & 1770 & 1920 & 8069     & 8096     & 3791 & -0.092$\pm$0.004 \\
$D_{35}(1930)$ & 1870 & 2100 & 9912     & 9881     & 5059 & -0.063$\pm$0.014 \\
$F_{37}(1950)$ & 1800 & 2000 &10623     &10552     & 4951 &  0.024$\pm$0.006 \\
\colrule
\end{tabular}
\end{table}
\newpage
\begin{table}[th]
\caption{Parameters for low-lying resonances of isospin 
         $1/2$ and $3/2$ in a fit unconstrained
         by dispersion relations. \label{tbl11}} 
\begin{tabular}{cccc}
\colrule
Resonance      & W$_R$ & $\Gamma$/2  & $\Gamma_{\pi N}/ \Gamma$ \\
               & (MeV) & (MeV)       &                          \\
\colrule
$P_{11}(1440)$ &1473.0 & 200.9       & 0.695                    \\
$D_{13}(1520)$ &1516.2 &  50.8       & 0.655                    \\
$S_{11}(1535)$ &1545.4 &  87.0       & 0.385                    \\
$S_{11}(1650)$ &1658.9 &  55.3       & 1.000                    \\
$D_{15}(1675)$ &1673.7 &  75.9       & 0.396                    \\
$F_{15}(1680)$ &1677.8 &  64.1       & 0.692                    \\
$P_{13}(1720)$ &1807.1 & 239         & 0.195                    \\
\colrule
$P_{33}(1232)$ &1233.3 &  59.3       & 1.000                    \\
$S_{31}(1620)$ &1614.0 &  70.1       & 0.316                    \\
$D_{33}(1700)$ &1684.9 & 168.6       & 0.151                    \\
$F_{35}(1905)$ &1861.3 & 159         & 0.119                    \\
\colrule
\end{tabular}
\end{table}
\begin{table}[th]
\caption{Comparison of $\chi^2$ for the FA02 (full data set) 
         and X370 (no charge-exchange data) solutions to 
         2.1~GeV ($\pi^-p\to\eta n$ data to 800~MeV).  Results 
         to 500~MeV have been listed in brackets, where one 
         observes little or no difference in the fit quality 
         between FA02 ands X370, even for the charge-exchange 
         database. ``Data" refers to the number of scattering 
         data used in the fit. \label{tbl12}}
\begin{tabular}{ccccc}
\colrule
Reaction          &  Observable       &  FA02        &  X370        & Data \\
\colrule
$\pi^+p\to\pi^+p$ & $d\sigma/d\Omega$ & 14574 (2043) & 14624 (2033) & 7246 (977)\\
                  &  $\sigma_{tot}$   &   257  (115) &   220  (115) &  105  (60)\\
                  &       P           &  6801  (533) &  6629  (536) & 3012 (495)\\
                  &       R           &    28   (20) &    29   (20) &   48  (26)\\
                  &       A           &    39   (24) &    42   (25) &   48  (26)\\
\colrule
$\pi^-p\to\pi^-p$ & $d\sigma/d\Omega$ & 13799 (1443) & 13760 (1453) & 7331 (773)\\
                  &  $\sigma_{tot}$   &   531  (111) &   445  (110) &  151  (59)\\
                  &       P           &  4441  (562) &  4501  (564) & 2045 (337)\\
                  &       R           &   107   (24) &   104   (23) &   61  (27)\\
                  &       A           &    49   (14) &    47   (14) &   60  (26)\\
\colrule
$\pi^-p\to\pi^0n$ & $d\sigma/d\Omega$ &  3180  (697) &  5093  (739) & 1333 (379)\\
                  &  $\sigma_{tot}$   &    24   (23) &    24   (23) &   34  (33)\\
                  &       P           &   933  (194) &  1175  (215) &  323 (159)\\
\colrule
$\pi^-p\to\eta n$ & $d\sigma/d\Omega$ &   373 &   376 &  138 \\
                  &  $\sigma_{tot}$   &    67 &    75 &   35 \\
\colrule
\end{tabular}
\end{table}
\newpage
\newpage
\newpage
\newpage


\begin{thebibliography}{99}
\bibitem{notation}In what follows, $\pm$ subscripts denote charge
              channels, superscripts denote isospin channels, $s$, 
              $t$, and $u$ are the usual Mandelstam variables, and
              $M$ is the nucleon mass.
\end{thebibliography}
\end{document}